\documentclass[aps,notitlepage,onecolumn]{revtex4-2}

\usepackage{graphicx}
\usepackage{chemformula}
\usepackage{lipsum}
\usepackage{graphicx}
\usepackage{subfigure}
\usepackage{color}
\usepackage{amsmath}
\usepackage[linktocpage,colorlinks=true,linkcolor=blue,citecolor=blue,urlcolor=blue,breaklinks=true]{hyperref}
\usepackage{verbatim}
\usepackage{breakcites}
\usepackage{wrapfig}
\usepackage{soul}
\usepackage{bbold}
\usepackage{gensymb}
\usepackage{textcomp}
\usepackage{amssymb}
\usepackage{mathtools}
\usepackage{xcolor}
\usepackage{verbatimbox}
\usepackage{multibib}
\usepackage{wasysym}
\usepackage{lineno}

\usepackage{xcolor,colortbl}
\usepackage{enumitem}
\usepackage{array}
\usepackage{cellspace}
    \newcolumntype{P}[1]{>{\centering\arraybackslash}p{#1}}

\DeclareMathOperator\erf{erf}

\begin{document}

\title{Light-induced phase separation with finite wavelength selection \\ in photophobic micro-algae}

\author{Isabelle Eisenmann$^1$, Alfredo L'Homme$^2$, Ali\'enor Lahlou$^{3,4}$, Sandrine Bujaldon$^5$, Thomas Le Saux$^3$, Benjamin Bailleul$^5$, Nicolas Desprat$^1$, Rapha\"el Jeanneret$^1$} 
\affiliation{ $^1$Laboratoire de Physique de l'Ecole normale sup\'erieure, ENS, Universit\'e PSL, CNRS, Sorbonne Universit\'e, Universit\'e Paris Cit\'e, F-75005 Paris, France\\ 
$^2$Institute for Biological and Medical Engineering, Schools of Engineering, Medicine and Biological Sciences, Pontificia Universidad Cat\'olica de Chile, Santiago, Chile\\ 
$^3$PASTEUR, Department of Chemistry, Ecole Normale Sup\'erieure, PSL University, Sorbonne University, CNRS, Paris, France\\
$^4$Sony Computer Science Laboratories, Paris, France\\
$^5$Laboratory of Chloroplast Biology and Light Sensing in Microalgae, Institut de Biologie Physico Chimique, CNRS, Sorbonne Universit\'e, Paris, 75005 France}	

\email[Correspondence: ]{raphael.jeanneret@phys.ens.fr, nicolas.desprat@phys.ens.fr}



\begin{abstract}
As for many motile micro-algae, the freshwater species \textit{Chlamydomonas reinhardtii} can detect light sources and adapt its motile behavior in response. Here, we show that suspensions of photophobic cells can be unstable to density fluctuations, as a consequence of shading interactions mediated by light absorption. In a circular illumination geometry this mechanism leads to the complete phase separation of the system into transient branching patterns, providing the first experimental evidence of finite wavelength selection in an active phase-separating system without birth and death processes. The finite wavelength selection, that can be captured in a simple drift-diffusion framework, is a consequence of a vision-based interaction length scale set by the illumination geometry and depends on global cell density, light intensity and medium viscosity. Finally we show that this active phase separation shields individual cells from the deleterious effects of high light intensity, demonstrating that phototaxis can efficiently contribute to photoprotection through collective behaviors on short timescales.

\end{abstract}

\maketitle

\subsection*{Introduction}
From penguins huddling on sea ice \cite{zitterbart_coordinated_2011} to bacteria forming fruiting bodies \cite{kaiser_coupling_2003} to microphase separation by molecular motors \cite{lemma_active_2022}, motility is essential to self-organization at all scales in living systems. Generically, systems driven out of equilibrium by self-propelled constituents can spontaneously phase-separate into a dense and a dilute phase, a mechanism called Motility-Induced Phase Separation (MIPS) \cite{Cates2015a,obyrne_introduction_2023}. MIPS relies either on particle velocity decreasing with particle density, stemming for instance from collisions, repulsive interactions or active biological responses such as quorum sensing, or on particle reorientation towards denser regions \cite{obyrne_lamellar_2020, zhang_active_2021}. 
At short time scales and for populations of constant size, MIPS generates patterns ranging from one to many clusters of variable size randomly distributed in space. Clusters then coarsen, unless other competing interactions such as hydrodynamics, chemotaxis or alignement arrest the phase separation \cite{matas-navarro_hydrodynamic_2014, zhao_chemotactic_2023, van_der_linden_interrupted_2019}. So far, no finite wavelength instabilities have been reported in active systems with motility-regulating interactions only. This marks a strong contrast with reaction-diffusion systems for instance, where patterns with a well-defined wavelength are commonly found \cite{kondo_reaction-diffusion_2010}.
Collections of micro-organisms constitute an interesting playground for investigating self-organization from relatively simple interaction rules between the agents \cite{ben-jacob_cooperative_2000}. 
Here we show that a dilute suspension of motile microalgae  \textit{Chlamydomonas reinhardtii} \cite{Harris2009, jeanneret_brief_2016} undergoing negative phototaxis (i.e. fleeing light sources) can phase separate into a dense liquid-like phase with selection of a finite wavelength within a few minutes of illumination. In our experimental circular geometry, the patterns take the shape of branched trees, with regularly spaced filaments converging at the center of the system. This instability is triggered by the interplay between negative phototaxis and light absorption by the cells: algae hide behind each other when fleeing multiple light sources, thus reinforcing initial density fluctuations. Our simple model suggests that the selection of a finite wavelength emerges from the non-locality of the shading interactions. Based on a drift-diffusion framework, this model captures essential experimental observations, like destabilization at critical cell density and light intensity and quantitative prediction of the wavelength of the instability. Finally we establish that algae gathered in the dense phase are collectively protected from the deleterious effects of light.

\begin{figure*}
	\centering
	\includegraphics[width=1\linewidth]{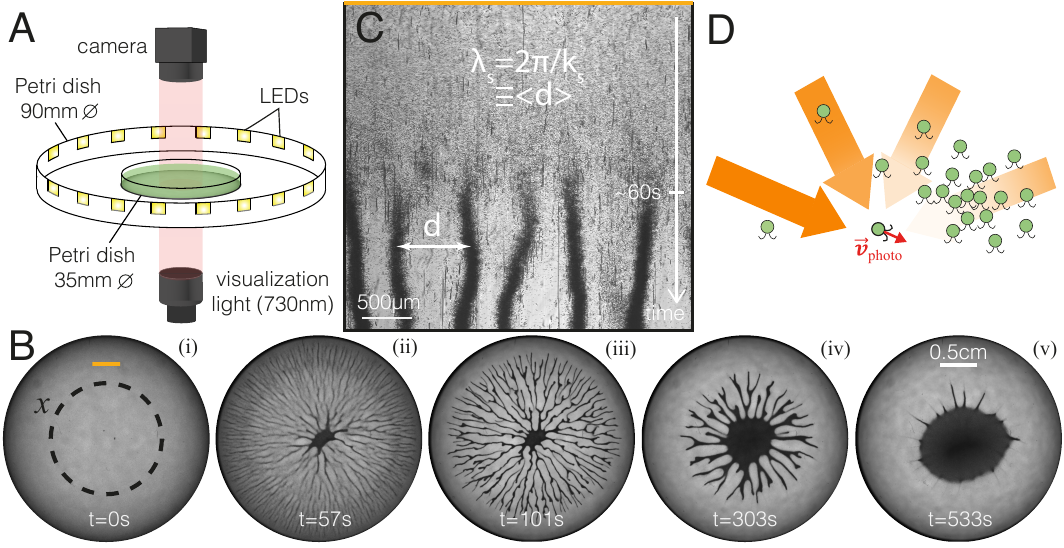}
	\caption{
		\textbf{A density instability from negative phototaxis and light absorption.}
		\textbf{A:} Schematics of the setup. A suspension of \textit{Chlamydomonas reinhardtii} in TAP medium is loaded in a Petri dish of diameter 3.5cm, surrounded by a ring of 16 LEDs glued to a larger Petri dish (9cm diameter). Visualization is done with a far red collimated light which does not interfere with phototaxis (730nm).
		\textbf{B:} Time-lapse of the light-induced phase separation (at cell concentration $\rho_0=2.8\times 10^6$cells/mL). Initially (before turning on the LED ring) the suspension is homogeneous (frame (i)). After switching on the lights, cells start migrating towards the center of the system, following a negative phototactic response. After one minute, dense filaments appear in the  whole Petri dish (frame (ii)). These then evolve into inter-connected dense branches that retract slowly towards the center (frames (iii) and (iv)). Finally a steady dense "drop" is found in the center of the dish (frame (v)). On panel B(i) we also show the definition of the coordinate $x$ used in the theoretical model. 
		\textbf{C:} Kymograph of the destabilization process along the yellow line in panel B(i), from microscopy recordings at $4\times$ magnification. The initial homogeneous suspension destabilizes into growing dynamical clusters that coarsen into short filaments and finally into dense branches separated by a well-defined wavelength $\lambda_s$.
		\textbf{D} {Schematics illustrating the mechanism of the instability: because of light absorption, a stochastic density fluctuation creates an inhomogeneous light environment for neighboring cells (orange arrows). These cells then flee from the most intense lights, towards the fluctuation, reinforcing it.}
		} 
	\label{fig:1}
\end{figure*}

\subsection*{Results}

\begin{figure*}
\centering
\includegraphics[width=0.9\linewidth]{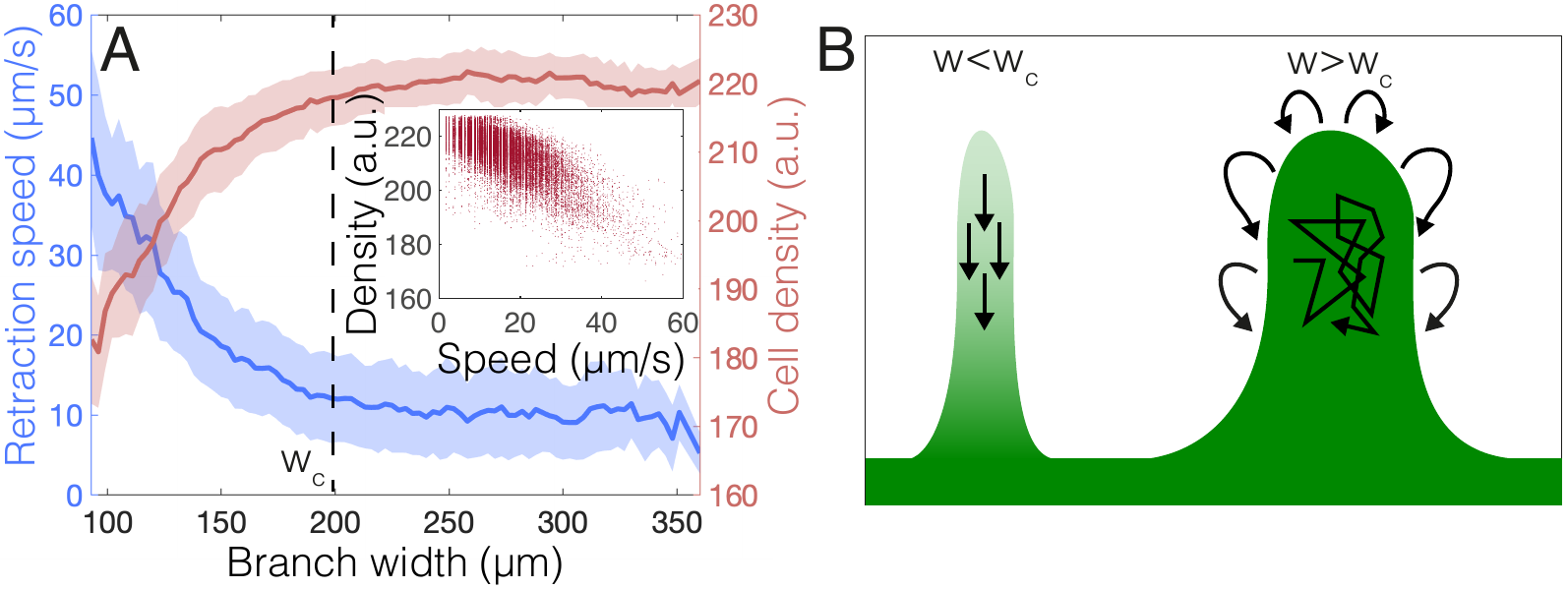}
\caption{\textbf{Dynamics of branch retraction.} \textbf{A:} The retraction speed of individual branches decreases with branch width until reaching a plateau at $\sim 10 {\rm \mu m/s}$ for $w>w_c\approx200 {\rm \mu m}$ (blue curve). This evolution mirrors that of the branch density (red curve) which saturates above $w_c$. The proxy for the cell density in the branches is simply the inverted gray level of our 8bits images. Retraction speed is computed over a time interval of 10s. These plots are obtained by discretizing the branch widths into bins $3 {\rm \mu m}$-wide and collecting data from four different replicas at cell concentration $2.8\times10^6$ cells/mL. Shaded area shows two standard deviations. Inset: Scatterplot of branch density versus retraction speed for all individual data points used to produce the main figure, and showing a clear linear correlation between the two quantities. \textbf{B:} When branches are thin enough ($w<w_c$), light goes through and cells proceed fleeing the lights, resulting in a net phototactic flux in the bulk of the branches and thus high retraction speeds (Fig. \ref{fig:2}B, Movie S5).  When branches are too wide ($w>w_c$), light intensity within the bulk of the branch becomes too low for the cells to perform phototaxis. The cell motion is then isotropic and dominated by collisions (Movie S4), producing no net phototactic flux inside wide branches. A downward flux is still present at the interface between the dense and dilute phase. 
}
\label{fig:2}
\end{figure*}

\subsubsection*{Photophobic micro-algal suspensions exhibited density instabilities and phase separation} 

To characterize the photophobic response of populations of \textit{C. reinhardtii} cells, we used a white LED strip surrounding cells suspended in liquid (in TAP medium) \cite{Harris2009}
and sitting in a cylindrical dish (Fig. \ref{fig:1}A and Materials and methods, cell concentration $\rho_0$ varied between $10^5$ and $5\times 10^6$ cells/mL). We monitored the suspension with a collimated far-red light (730nm) that doesn't interfere with phototaxis. In our conditions cells were seen to only perform negative phototaxis in the range of light intensity we used (spatial average intensity per LED within the inner Petri dish estimated to be $I_{\rm LED}\ge 4{\rm mW/m^2}$). 
We homogenized the suspension and then switched on the LED lights (Fig. \ref{fig:1}B(i)), triggering individual cells to migrate towards the center of the dish. 
After $\sim 30$s denser areas started forming throughout the Petri dish (Movie S1). These denser areas then evolved into long and connected "branches" with sharp edges that converge in the center of the dish (Fig. \ref{fig:1}B(ii)-(iii)), thus creating a single dense micro-algal phase that spans the whole system. Seen under a microscope (Movie S2 and Fig. \ref{fig:1}C), after phase separation the emerging "liquid" phase was characterized by
seemingly isotropic motion of the cells in its bulk. The average inter-branch distance $\lambda_s\equiv\langle d\rangle $ (Fig. \ref{fig:1}C) was measured to be independent of radial position during early pattern retraction (SI section S1, Fig. S1), and will be used in the following to characterize the density instability. The final retraction of the pattern involved branches emptying in the vertices, leading to increasing branch width and inter-branch distance with time (Movie S1), until all branches had emptied into the cylindrical "drop" at the center of the system (Fig. \ref{fig:1}B(iv)-(v)).

No adhesion was involved in the density instability and phase separation, as evidenced by the fact that at any moment if the light sources were switched off the pattern immediately started to blur (Movie S3) due to the diffusive run-and-tumble motion of the algae in the dark (diffusion in the dark measured to be $ D_0=1.7\pm 0.4 \times10^{4}{\rm \mu m^2/s}$, SI section S2, Fig. S2) \cite{polin_chlamydomonas_2009}. 
Moreover, the formation of the pattern was not influenced by bioconvective effects \cite{Bees2020}, since cells also migrated towards the bottom of the dish given the geometry of the setup (Fig. S3), creating patterns effectively confined within $\sim 50-60 {\rm \mu m}$ at the bottom of the dish (SI section S3 and Fig. S4). We estimated that the density of the dense phase was $\rho_{\rm drop}\approx3\times 10^9$ cells/mL ($\sim 50\%$ volume fraction), three orders of magnitude larger than the initial cell concentration.
Finally, the number of branches and their location were not correlated with the discreteness of the light sources, as shown by e.g. superimposing the patterns obtained in different replica (Fig. S5A). This was true as long as the inter-LED distance remained short such that the light cones overlapped within the inner Petri dish (SI section S4). On the contrary, if the stimulating lights were too distant from each other, cells only accumulated in the angular regions not facing the lights (Fig. S5B).

\subsubsection*{Light absorption by the cells was essential to understand their interactions}

Once formed, the branching pattern retracted relatively slowly compared to the typical swimming speed of individual cells ($\sim 10 {\rm \mu m/s}$, Fig. S1C, versus $\sim 100 {\rm \mu m/s}$ respectively). Looking at individual branches, we noticed that thin branches retracted much faster than wide ones (Movie S1). This was confirmed quantitatively by tracking the tip of individual branches. We showed that on average the speed of tip retraction decreased with branch width until reaching a plateau for $w>w_c\sim150-200 {\rm \mu m}$ (Fig. \ref{fig:2}A, blue curve). On the contrary, on average the cell concentration in the branches increased with branch width until plateauing when the width exceeds $w_c$ (Fig. \ref{fig:2}A, red curve). 
The anti-correlation between branch retraction speed and cell density (Fig. \ref{fig:2}A-inset) could be understood by considering light absorption by the cells. In the dense phase, the characteristic length scale for light absorption by the cell assembly was estimated to be $L_{\rm abs}\approx 10 {\rm \mu m}$ ($\sim$ two cells width, see SI section S5 for details). When branches were a few tens of $L_{\rm abs}$ wide, light intensity within the bulk of the branch became too low for the cells to perform phototaxis (intensity from a single LED estimated to be $I_{\rm bulk}\approx0.1 {\rm mW/m^2}$ inside a branch of width $w_c$). The cell motion was then isotropic and dominated by collisions (Movie S4), producing no net phototactic flux inside wide branches. However, a downward flux was still present at the interface between the dense and dilute phase (Fig. \ref{fig:2}B).
Such a mechanism led to a net retraction speed of the branches mostly independent of their width. 
However once branches became thin enough, light penetrated into the branch and all cells proceeded to flee the light source, resulting in a net phototactic flux in the bulk of the branches and thus higher retraction speeds (Fig. \ref{fig:2}B, Movie S5).  
The appearance of the first regime where cell motion was radically modified (isotropic and dominated by collisions) constituted a strong support for classifying the dense phase as a liquid phase, further supporting the qualification of \textit{phase separation} for the phenomenon we describe. 

\begin{figure*}
\centering
\includegraphics[width=1\linewidth]{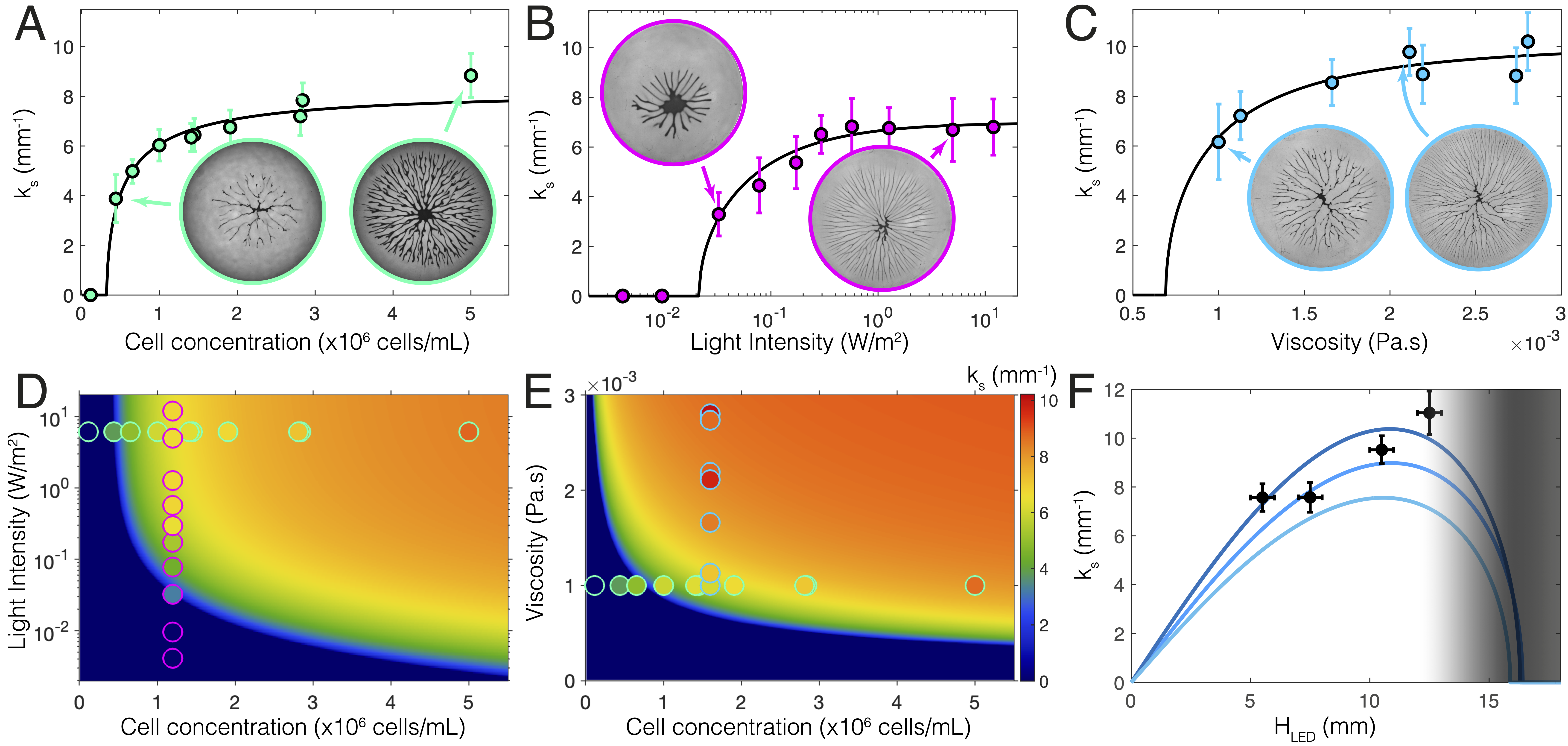}
\caption{\textbf{Quantification of the density instability and comparison with the model.} \textbf{A-C:} Experimentally measured wavenumber of the instability $k_s$ as a function of the experimental control parameters, A. cell density, B. average light intensity per LED and C. medium viscosity. Error-bars represent standard deviations of at least 3 replicas. Black solid lines correspond to the fitted analytical expression $k_s^{\rm (th)}$ (Eq. \ref{ks_exp}). Insets: snapshots of the system at the time at which $k_s$ is evaluated (section S1) for different conditions.  \textbf{D-E:} Predicted phase diagram of the system where, for each set of variables $(\rho_0,I_{\rm LED})$ and $(\rho_0,\eta)$, we take the average value of the fitting parameters $a$ and $\Delta x$. \textbf{F:} Experimentally measured wavenumber of the instability $k_s$ as a function of the height of the LED ring in the setup $H_{\rm LED}$ at fixed cell concentration $\rho_0=2.1\times 10^6$cells/mL and fixed light intensity $I_{\rm LED}=6.2 {\rm W/m^2}$ (black circles). The blue curves are theoretical predictions based on the expression of $k_s^{\rm (th)}$ with $\Delta x=\Delta x_{\rm geo}(H_{\rm LED})$ using the previously measured average values of the parameters involved (light blue) or when considering one standard deviation around the mean of each parameter (dark blue and cyan). The gray shaded region approximately shows where the simple 1D approach is expected to fail to capture the behavior of the system.
}
\label{fig:3}
\end{figure*}

Light absorption by branches in turn influenced the motion of cells in neighbouring branches. As shown in Movie S6 thin branches could be locally attracted by wider ones and flow into them, deviating from pure radial retraction. Cells swimming close to a wide branch perceive an asymmetric light environment due to light absorption by this very dense area nearby. They then started biasing their own motion to flee from the locally most intense light, i.e. towards the nearby branch. This result showed that cells are sensitive to even minute changes in the asymmetry of light stimulation, that can be mediated by local heterogeneities in cell density.

\subsubsection*{A density instability arising from negative phototaxis and light absorption}

Based on the above observations, we proposed a simple mechanism for explaining the initial destabilization and phase separation of the photophobic suspension. When turning on the lights initially, cells movement oriented radially towards the center away from the light sources (movie S2, after a short transient positive phototaxis \cite{uhl_adaptation_1990,takahashi_photosynthesis_1993}). 
At this stage, cells mainly detected light coming from their side since the eyespot - the light detector located on the cell's equator - spins along the radial swimming direction \cite{foster_light_1980, schaller_how_1997, bennett_steering_2015, leptos_phototaxis_2023}.
Therefore, any spontaneous positive fluctuation of the local cell density along the azimuthal direction could modify the phototactic response of the cells lying within the shadow of the fluctuation, by biasing motion towards the denser region (see schematics in Fig. \ref{fig:1}D).
The initial stochastic density fluctuation could be therefore be amplified, leading to the destabilization of the suspension and the phase separation of the system. The destabilization process was very dynamical, since initial density fluctuations appear everywhere in a completely stochastic manner, leading to the erratic growth of clusters, that quickly form short filaments which themselves evolve into longer inter-connected branches (Movie S2). 

Following this proposed mechanism, we now describe a simple phenomenological model to account for the density instability and quantitatively compare with our experimental results. We consider a 1D continuum description of the cell density field $\rho(x,t)$ along the azimuthal direction $x$ (see Fig. 1B for the definition). The one-dimensional approach is justified in SI section S6.1 and further discussed in section S12. We model the evolution of the density field with a drift-diffusion equation (SI section 6.2)
\begin{align}
\label{drift_diffusion}
\frac{\partial \rho}{\partial t}  &= D \frac{\partial^2 \rho}{\partial x^2} - \frac{\partial \rho v_{\rm photo}}{\partial x}\\
\label{vphoto}
v_{\rm photo}(x,t) &= \alpha\Big[\int_x^{x+\Delta x} \rho(x',t)dx' - \int_{x-\Delta x}^{x} \rho(x',t)dx'\Big]
\end{align}
where the drift term $v_{\rm photo}$ describes the phototactic behavior of the cells that tend to bias azimuthally their motion towards region of higher densities, as they migrate radially towards the center of the dish. We assume that the cells are able to detect left-right asymmetries in cell density with a sensitivity $\alpha$, only over a finite interaction distance $\Delta x$ as is usually the case for vision-based interactions between agents \cite{ballerini_interaction_2008, ginelli_intermittent_2015, filella_model_2018}. Noise in the ability of the cells to follow specific direction is modeled through a constant diffusion $D$. 

By performing a linear stability analysis of these equations (SI section S6.3) we showed that the homogeneous state $\rho(x,t)=\rho_0$ is unstable as soon as the parameter $\beta=\alpha\rho_0/D$ is larger than a critical value $\beta_c=1/\Delta x^2$. In such case the selected mode is predicted to be
\begin{align}
\label{Most_unstable}
k_{s}^{\rm (th)}=\frac{1}{\Delta x}{\rm sinc}_{[0;\pi]}^{-1}\Big(\frac{1}{\beta \Delta x^2}\Big)
\end{align}
where ${\rm sinc}_{[0;\pi]}^{-1}(y)$ $(y\in [0,1])$ is the inverse function of the sine cardinal ${\rm sinc}(x)$, $x \in [0, \pi]$. When $\beta$ becomes much larger than $\beta_c$, $k_s^{\rm (th)}$ saturates at a value $k_{s, {\rm max}}^{\rm (th)}=\pi/\Delta x$, which is set by the interaction length $\Delta x$. Put simply, if two density peaks develop closer than twice the interaction distance, they merge, leaving large peaks further apart than $2\Delta x$ on average. 

This model can be compared to the experimental quantification of the selected wavenumber at the onset of pattern formation when varying different control parameters, namely the initial cell concentration $\rho_0$, the average light intensity per LED in the Petri dish $I_{\rm LED}$ (see definition in SI section S7) and the viscosity of the medium $\eta$ (by adding Methyl Cellulose to the TAP medium, SI section S8). The wavenumber is here defined as $k_s=2\pi/\lambda_s$ with $\lambda_s$ being the average inter-branch distance (Fig. \ref{fig:1}C, SI section S1).  Each parameter was varied while keeping the other two fixed. For each dataset, the wavenumber $k_s$ increased as the experimental parameter was increased (Fig. \ref{fig:3}A-C, circles, and movies S7-9), meaning that branches got closer to each others. When the cell density was fixed (i.e. when varying $I_{\rm LED}$ or $\eta$) this also implied that branches got thinner (Fig. \ref{fig:3}B-C - Insets and Fig. S10). Moreover, the experimental data points seemed to saturate at a similar value $k_{s,{\rm max}}\sim 7$-$10 {\rm mm^{-1}}$ (corresponding to a minimal average inter-branch distance $\lambda_{s,{\rm min}}\sim 600$-$900 {\rm \mu m}$). Finally, at low cell density ($\rho_0<0.3\times 10^6 {\rm cells/mL}$) or light intensity ($I_{\rm LED}<3\times 10^{-2} {\rm W/m^2}$), the system did not destabilize along the azimuthal direction, but cells still performed negative phototaxis by migrating towards the center of the dish and accumulate there (movie S7 and S8, see SI section S9 for a more detailed discussion of these experiments).

To proceed to a comparison with the model, we had to make hypothesis as to how the parameters $D$ and $\alpha$ (and therefore $\beta$) depend on our experimental parameters $\eta$ and $I_{\rm LED}$ (SI section S10.1). First, the diffusion coefficient $D$ was taken to be the run-and-tumble diffusivity in the dark ($\propto U^2 \tau$ with $U$ the speed of the algae, $\tau$ the run time)
and therefore assumed to be written $D(\eta)=D_0 \eta_0^2/\eta^2$ (with $D_0$ the diffusivity measured in TAP medium of viscosity $\eta_0=1 {\rm mPa.s}$, see SI section S11 for a discussion on this diffusivity $D$).
Second, the parameter $\alpha$ was modeled according to previous experiments done on channelrhodopsin molecules in different in vitro and in vivo systems \cite{hartmann_hartz_photoreceptor_1992, rorsman_defining_2018, wang_high-speed_2007}, where their light intensity response was shown to follow a Hill function with an exponent $p$ robustly extracted to be $\approx0.7$. Therefore we wrote $\alpha=aI_{\rm LED}^p/(K^p+I_{\rm LED}^p)$, with $p=0.7$. 
Overall this led to the following prediction for the experimental evolution of the selected mode $k_s$
\begin{align}
\label{ks_exp}
k_{s} &= k_{s}^{\rm (th)}(\rho_0,\eta, I_{LED})   \nonumber \\
&= \frac{1}{\Delta x}{\rm sinc}_{[0;\pi]}^{-1}\Big(\frac{D_0}{a\rho_0\Delta x^2}\frac{\eta_0^2}{\eta^2}\frac{K^p+I_{\rm LED}^p}{I_{\rm LED}^p}\Big)
\end{align}

This model reproduced very well our experimental observations (black curves in Fig. \ref{fig:3}A-C) and returned fitting parameters $\Delta x\approx 340 {\rm \mu m}$ and $a\approx 4\times 10^{-13} {\rm m^3/s}$ consistent between the three sets of independent experiments (SI section S10.2), demonstrating that our simple 1D approach well captured the physics of the phenomenon. This allowed us to draw the expected phase diagram of the system obtained from the mean values of the fitting parameters and to compare it to the experimental points (Fig. \ref{fig:3}E-F). 
The interaction length $\Delta x$ extracted from the fit to the model was large, about forty times the size of an alga, and the shading interactions appearing between the cells were therefore highly non-local. We noticed that an interaction length of $340 {\rm \mu m}$ was compatible with the size of the shadows created by the branches in our illumination geometry. For a branch height of $h_{\rm branch}\approx 60 {\rm \mu m}$ (SI section S3) and a typical light angle $\gamma\approx 9.5^{\circ}$ (Fig. S3), the typical shadow was indeed $\Delta x_{\rm geo}=h_{\rm branch}/\tan(\gamma)=h_{\rm branch}R_{\rm LED}/H_{\rm LED}\approx360 {\rm \mu m}$  (see Fig. S12 and SI section S12 for details). Changing the illumination angle $\gamma$ should therefore modify the selected wavenumber. 
We tested this hypothesis by changing the height of the LED ring between $H_{\rm LED}\approx5.5$mm and 12.5mm at fixed cell concentration, light intensity and medium viscosity and obtained wavenumbers $k_s$ consistent with our model when considering $\Delta x=\Delta x_{\rm geo}(H_{\rm LED})$ (Fig. \ref{fig:3}F and movie S12) without fitting parameters (SI section S12). While providing a possible explanation for the origin of the interaction length $\Delta x$, this analysis also showed the predictive ability of our simple theoretical approach.

\subsubsection*{In the dense phase cells were protected against photodamages}

\begin{figure}
	\centering
	\includegraphics[width=0.6\linewidth]{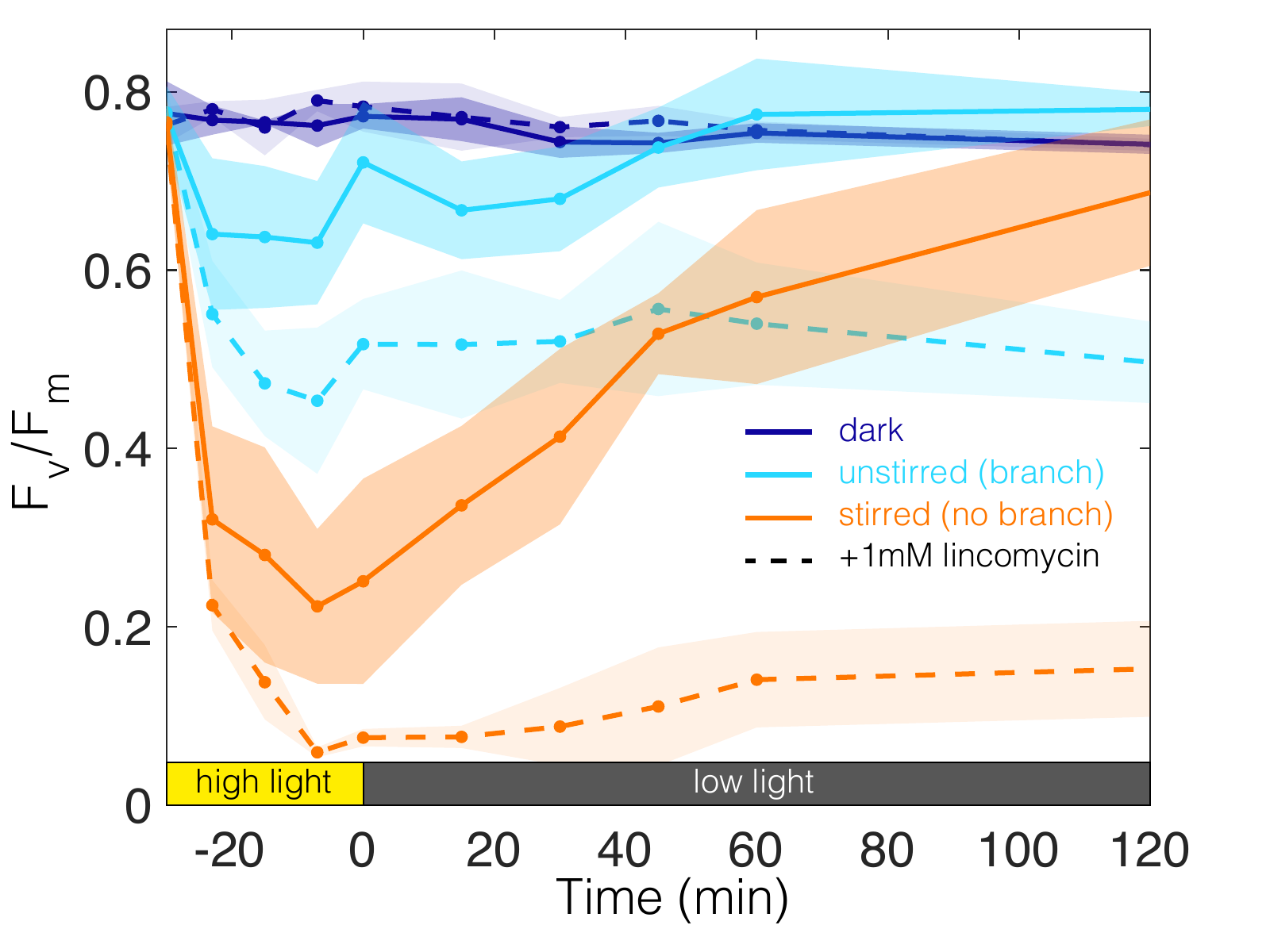}
	\caption{\textbf{Quantification of photoinhibition through fluorescence measurements.} To quantify photoinhibition, cells are exposed to 30min blue light at high intensity ($I_{\rm tot}=78{\rm W/m^2}=303 {\rm \mu mol/m^2/s}$ in the instability setup, either letting cells phase-separate (light blue curves) or preventing them from forming a dense phase by gently shaking the setup on an orbital shaker at 100rpm (orange curves). A control is kept in the dark (dark blue curves). During the high-light phase, photoinhibition is much larger when cells could not develop a dense phase.  Both conditions lead to the relaxation of $F_v/F_m$ to its initial value in the next two hours of low light conditions (solid lines). Adding the chloroplast inhibitor lincomycin prevents such relaxation (dashed lines) confirming that the drop in $F_v/F_m$ is indeed induced by the high-light treatment. Shaded areas correspond to one standard deviation of 3 biological replicas.
	}
	\label{fig:4}
\end{figure}

Between the initial dilute and homogeneous state and the emerging dense phase, the characteristic length scale for light absorption decreased from a few centimeters to a couple of cells widths. We then wondered whether the crowded environment of the dense phase was beneficial for the algae with regards to the harmful effects of excess light on the photosynthetic machinery, the main effect being the photoinactivation of Photosystem II (PSII). Such harmful effects are routinely quantified via the maximal photochemical efficiency of PSII, or $F_v/F_m$ ratio, through chlorophyll fluorescence measurements \cite{baker_chlorophyll_2008}.
Our results (Fig. \ref{fig:4}, SI section S13 and Materials and Methods) showed that the decrease of the ratio $F_v/F_m$ was much smaller in the presence of branching patterns compared to a situation where cells were continuously stirred (respectively solid blue and orange lines in Fig. \ref{fig:4}). We also showed that the decrease of Fv/Fm was amplified when the repair of damaged PSII was prevented by the chloroplast translation inhibitor lincomycin (dashed blue and orange lines in Fig. \ref{fig:4}). This indicated that the emergent light-induced phase-separation of the suspension provided a collective function to the cell population by effectively protecting it from the harmful lights.
Such collective protection, reminiscent of chloroplast reorganization in plants \cite{schramma_chloroplasts_2023}, builds up quickly ($\sim$ minute) and therefore complements well other individual photo-protective mechanisms that require longer light exposure ($\sim$ minutes to hours) for synthesizing effector proteins \cite{erickson_light_2015}.

\subsection*{Discussion}

In our system, suspensions of photophobic micro-algae destabilize and phase separate into ordered branching patterns characterized by a finite wavelength. Our phenomenological model demonstrates that the non-locality of the shading interactions between the cells is essential to explain the selected patterns. This shows that non-locality is an important feature of tactic interactions that could play a role in other experimental situations, e.g. to quantitatively explain the chemotactic patterns with $\sim$ mm characteristic length scales formed by swarming and dividing \textit{E. coli} cells in Burdene and Berg's experiments \cite{budrene_complex_1991}. Another specificity of the shading interactions is their anisotropy and non-reciprocity induced by the eyespot location that captures light within a given cone that laterally scans the light environment as the cell spins along its swimming direction \cite{foster_light_1980, schaller_how_1997, bennett_steering_2015, leptos_phototaxis_2023}. Here, even though the cells move towards the center of the setup fleeing the light sources closest to them, the destabilization occurs in the azimuthal direction. Fluctuations in the azimuthal direction indicate the direction of the denser regions, towards which the cells rectify their motion. 
In addition, the quantification of the intensity fields using a recently developed fluorescence technique \cite{Lahlou23} (Fig. S13 and SI section S14) suggests that in our setup, cells do not follow the light intensity gradients, as had been found in other experimental situations \cite{arrieta_phototaxis_2017, ramamonjy_nonlinear_2022}. Such observations raise questions regarding the way these cells actually use light as a guide to navigate, in particular when facing multiple light sources. Our understanding of phototaxis would therefore gain in having dedicated experiments to better link microscopic swimming properties to coarse-grained continuum equations at the population level. Finally, the phenomenon at play, which leads to the formation of a dense phase with large volume fraction $\sim 50 \%$, could benefit the microalgal industry in order to pre-concentrate the biomass before harvesting, that currently represents a real bottleneck in the use of these organisms for clean energy, drugs or food production \cite{koyande_microalgae_2019,ferreira_mota_biodiesel_2022}.

\bibliography{Collective_photoprotection}

\begin{thebibliography}{10}

\bibitem{zitterbart_coordinated_2011}
D.~P. Zitterbart, B.~Wienecke, J.~P. Butler, and B.~Fabry,
\newblock {\em Coordinated {Movements} {Prevent} {Jamming} in an {Emperor}
  {Penguin} {Huddle}}, PLOS ONE {\bf 6}, e20260 (2011).

\bibitem{kaiser_coupling_2003}
D.~Kaiser,
\newblock {\em Coupling cell movement to multicellular development in
  myxobacteria}, Nat. Rev. Microbiol. {\bf 1}, 45 (2003).

\bibitem{lemma_active_2022}
B.~Lemma, N.~P. Mitchell, R.~Subramanian, D.~J. Needleman, and Z.~Dogic,
\newblock {\em Active {Microphase} {Separation} in {Mixtures} of {Microtubules}
  and {Tip}-{Accumulating} {Molecular} {Motors}}, Phys. Rev. X {\bf 12}, 031006
  (2022).

\bibitem{Cates2015a}
M.~E. Cates and J.~Tailleur,
\newblock {\em Motility-{Induced} {Phase} {Separation}}, Annu. Rev. Condens.
  Matter Phys. {\bf 6}, 219 (2015).

\bibitem{obyrne_introduction_2023}
J.~O'Byrne, A.~Solon, J.~Tailleur, and Y.~Zhao,
\newblock {\em {An Introduction to Motility-induced Phase Separation}},
\newblock in {\em {Out-of-equilibrium Soft Matter}}, The Royal Society of
  Chemistry, 2023.

\bibitem{obyrne_lamellar_2020}
J.~O’Byrne and J.~Tailleur,
\newblock {\em Lamellar to {Micellar} {Phases} and {Beyond}: {When} {Tactic}
  {Active} {Systems} {Admit} {Free} {Energy} {Functionals}}, Phys. Rev. Lett.
  {\bf 125}, 208003 (2020).

\bibitem{zhang_active_2021}
J.~Zhang, R.~Alert, J.~Yan, N.~S. Wingreen, and S.~Granick,
\newblock {\em Active phase separation by turning towards regions of higher
  density}, Nat. Phys. {\bf 17}, 961 (2021).

\bibitem{matas-navarro_hydrodynamic_2014}
R.~Matas-Navarro, R.~Golestanian, T.~B. Liverpool, and S.~M. Fielding,
\newblock {\em Hydrodynamic suppression of phase separation in active
  suspensions}, Phys. Rev. E {\bf 90}, 032304 (2014).

\bibitem{zhao_chemotactic_2023}
H.~Zhao, A.~Košmrlj, and S.~S. Datta,
\newblock {\em Chemotactic {Motility}-{Induced} {Phase} {Separation}}, Phys.
  Rev. Lett. {\bf 131}, 118301 (2023).

\bibitem{van_der_linden_interrupted_2019}
M.~N. Van Der~Linden, L.~C. Alexander, D.~G. Aarts, and O.~Dauchot,
\newblock {\em Interrupted {Motility} {Induced} {Phase} {Separation} in
  {Aligning} {Active} {Colloids}}, Phys. Rev. Lett. {\bf 123}, 098001 (2019).

\bibitem{kondo_reaction-diffusion_2010}
S.~Kondo and T.~Miura,
\newblock {\em Reaction-{Diffusion} {Model} as a {Framework} for
  {Understanding} {Biological} {Pattern} {Formation}}, Science {\bf 329}, 1616
  (2010).

\bibitem{ben-jacob_cooperative_2000}
E.~Ben-Jacob, I.~Cohen, and H.~Levine,
\newblock {\em Cooperative self-organization of microorganisms}, Adv. Phys.
  {\bf 49}, 395 (2000).

\bibitem{Harris2009}
E.~S. Harris,
\newblock {\em The {Chlamydomonas} sourcebook} (Elsevier, 2009).

\bibitem{jeanneret_brief_2016}
R.~Jeanneret, M.~Contino, and M.~Polin,
\newblock {\em A brief introduction to the model microswimmer {Chlamydomonas}
  reinhardtii}, Eur. Phys. J. Spec. Top. {\bf 225}, 2141 (2016).

\bibitem{polin_chlamydomonas_2009}
M.~Polin, I.~Tuval, K.~Drescher, J.~P. Gollub, and R.~E. Goldstein,
\newblock {\em \textit{{Chlamydomonas}} {Swims} with {Two} “{Gears}” in a
  {Eukaryotic} {Version} of {Run}-and-{Tumble} {Locomotion}}, Science {\bf
  325}, 487 (2009).

\bibitem{Bees2020}
M.~A. Bees,
\newblock {\em Advances in {Bioconvection}}, Annu. Rev. Fluid Mech. {\bf 52},
  449 (2020).

\bibitem{uhl_adaptation_1990}
R.~Uhl and P.~Hegemann,
\newblock {\em Adaptation of {Chlamydomonas} phototaxis: {I}. {A}
  light-scattering apparatus for measuring the phototactic rate of
  microorganisms with high time resolution}, Cell Motil. Cytoskel. {\bf 15},
  230 (1990).

\bibitem{takahashi_photosynthesis_1993}
T.~Takahashi and M.~Watanabe,
\newblock {\em Photosynthesis modulates the sign of phototaxis of wild-type
  \textit{{Chlamydomonas} reinhardtii}: {Effects} of red background
  illumination and 3-(3',4'-dichlorophenyl)-1,1-dimethylurea}, FEBS Letters
  {\bf 336}, 516 (1993).

\bibitem{foster_light_1980}
K.~W. Foster and R.~D. Smyth,
\newblock {\em Light {Antennas} in phototactic algae}, Microbiol. Rev. {\bf
  44}, 572 (1980).

\bibitem{schaller_how_1997}
K.~Schaller, R.~David, and R.~Uhl,
\newblock {\em How {Chlamydomonas} keeps track of the light once it has reached
  the right phototactic orientation}, Biophys. J. {\bf 73}, 1562 (1997).

\bibitem{bennett_steering_2015}
R.~R. Bennett and R.~Golestanian,
\newblock {\em A steering mechanism for phototaxis in
  \textit{{Chlamydomonas}}}, J. R. Soc. Interface. {\bf 12}, 20141164 (2015).

\bibitem{leptos_phototaxis_2023}
K.~C. Leptos, M.~Chioccioli, S.~Furlan, A.~I. Pesci, and R.~E. Goldstein,
\newblock {\em Phototaxis of {Chlamydomonas} arises from a tuned adaptive
  photoresponse shared with multicellular {Volvocine} green algae}, Phys. Rev.
  E {\bf 107}, 014404 (2023).

\bibitem{ballerini_interaction_2008}
M.~Ballerini {\em et~al.},
\newblock {\em Interaction ruling animal collective behavior depends on
  topological rather than metric distance: evidence from a field study.}, Proc.
  Natl. Acad. Sci. U.S.A {\bf 105}, 1232 (2008).

\bibitem{ginelli_intermittent_2015}
F.~Ginelli {\em et~al.},
\newblock {\em Intermittent collective dynamics emerge from conflicting
  imperatives in sheep herds}, Proc. Natl. Acad. Sci. U.S.A. {\bf 112}, 12729
  (2015).

\bibitem{filella_model_2018}
A.~Filella, F.~Nadal, C.~Sire, E.~Kanso, and C.~Eloy,
\newblock {\em Model of {Collective} {Fish} {Behavior} with {Hydrodynamic}
  {Interactions}}, Phys. Rev. Lett. {\bf 120}, 35 (2018).

\bibitem{hartmann_hartz_photoreceptor_1992}
H.~Hartz, C.~Nonnengässer, and P.~Hegemann,
\newblock {\em The photoreceptor current of the green alga
  \textit{{Chlamydomonas}}}, Phil. Trans. R. Soc. Lond. B {\bf 338}, 39 (1992).

\bibitem{rorsman_defining_2018}
N.~J.~G. Rorsman, C.~M. Ta, H.~Garnett, P.~Swietach, and P.~Tammaro,
\newblock {\em Defining the ionic mechanisms of optogenetic control of vascular
  tone by channelrhodopsin-2: {Channelrhodopsin}-2 in vascular optogenetics},
  Br. J. Pharmacol. {\bf 175}, 2028 (2018).

\bibitem{wang_high-speed_2007}
H.~Wang {\em et~al.},
\newblock {\em High-speed mapping of synaptic connectivity using
  photostimulation in {Channelrhodopsin}-2 transgenic mice}, Proc. Natl. Acad.
  Sci. U.S.A. {\bf 104}, 8143 (2007).

\bibitem{baker_chlorophyll_2008}
N.~R. Baker,
\newblock {\em Chlorophyll {Fluorescence}: {A} {Probe} of {Photosynthesis} {In}
  {Vivo}}, Annu. Rev. Plant Biol. {\bf 59}, 89 (2008).

\bibitem{schramma_chloroplasts_2023}
N.~Schramma, C.~Perugachi~Israëls, and M.~Jalaal,
\newblock {\em Chloroplasts in plant cells show active glassy behavior under
  low-light conditions}, Proc. Natl. Acad. Sci. U.S.A {\bf 120}, e2216497120
  (2023).

\bibitem{erickson_light_2015}
E.~Erickson, S.~Wakao, and K.~K. Niyogi,
\newblock {\em Light stress and photoprotection in \textit{{Chlamydomonas}
  reinhardtii}}, Plant J. {\bf 82}, 449 (2015).

\bibitem{budrene_complex_1991}
E.~O. Budrene and H.~C. Berg,
\newblock {\em Complex patterns formed by motile cells of {Escherichia} coli},
  Nature {\bf 349}, 630 (1991).

\bibitem{Lahlou23}
A.~Lahlou {\em et~al.},
\newblock {\em Fluorescence to measure light intensity}, Nat. Methods {\bf 20},
  1930 (2023).

\bibitem{arrieta_phototaxis_2017}
J.~Arrieta, A.~Barreira, M.~Chioccioli, M.~Polin, and I.~Tuval,
\newblock {\em Phototaxis beyond turning: persistent accumulation and response
  acclimation of the microalga {Chlamydomonas} reinhardtii}, Sci. Rep. {\bf 7},
  3447 (2017).

\bibitem{ramamonjy_nonlinear_2022}
A.~Ramamonjy, J.~Dervaux, and P.~Brunet,
\newblock {\em Nonlinear {Phototaxis} and {Instabilities} in {Suspensions} of
  {Light}-{Seeking} {Algae}}, Phys. Rev. Lett. {\bf 128}, 258101 (2022).

\bibitem{koyande_microalgae_2019}
A.~K. Koyande {\em et~al.},
\newblock {\em Microalgae: {A} potential alternative to health supplementation
  for humans}, Food Sci. Hum. Wellness {\bf 8}, 16 (2019).

\bibitem{ferreira_mota_biodiesel_2022}
G.~Ferreira~Mota {\em et~al.},
\newblock {\em Biodiesel production from microalgae using lipase-based
  catalysts: {Current} challenges and prospects}, Algal Res. {\bf 62}, 102616
  (2022).

\bibitem{berne_peculiar_2018}
N.~Berne, T.~Fabryova, B.~Istaz, P.~Cardol, and B.~Bailleul,
\newblock {\em The peculiar {NPQ} regulation in the stramenopile {Phaeomonas}
  sp. challenges the xanthophyll cycle dogma}, Biochim. Biophys. Acta -
  Bioenerg. {\bf 1859}, 491 (2018).

\bibitem{de_mooij_impact_2016}
T.~De~Mooij, G.~De~Vries, C.~Latsos, R.~H. Wijffels, and M.~Janssen,
\newblock {\em Impact of light color on photobioreactor productivity}, Algal
  Res. {\bf 15}, 32 (2016).

\bibitem{bright_two-dimensional_1987}
D.~S. Bright and E.~B. Steel,
\newblock {\em Two-dimensional top hat filter for extracting spots and spheres
  from digital images}, J. Microsc. {\bf 146}, 191 (1987).

\bibitem{Vincent1991}
L.~Vincent and P.~Soille,
\newblock {\em Watersheds in digital spaces: an efficient algorithm based on
  immersion simulations}, IEEE Trans. Pattern Anal. Mach. Intell. {\bf 13}, 583
  (1991).

\bibitem{giometto_generalized_2015}
A.~Giometto, F.~Altermatt, A.~Maritan, R.~Stocker, and A.~Rinaldo,
\newblock {\em Generalized receptor law governs phototaxis in the phytoplankton
  \textit{{Euglena} gracilis}}, Proc. Natl. Acad. Sci. U.S.A. {\bf 112}, 7045
  (2015).

\bibitem{leroyer_driftdiffusion_2010}
Y.~Leroyer and A.~Würger,
\newblock {\em Drift–diffusion kinetics of a confined colloid}, J. Phys.:
  Condens. Matter {\bf 22}, 195104 (2010).

\bibitem{anguige_one-dimensional_2009}
K.~Anguige and C.~Schmeiser,
\newblock {\em A one-dimensional model of cell diffusion and aggregation,
  incorporating volume filling and cell-to-cell adhesion}, J. Math. Biol. {\bf
  58}, 395 (2009).

\bibitem{armstrong_continuum_2006-1}
N.~J. Armstrong, K.~J. Painter, and J.~A. Sherratt,
\newblock {\em A continuum approach to modelling cell–cell adhesion}, Journal
  of Theoretical Biology {\bf 243}, 98 (2006).

\bibitem{strogatz2018nonlinear}
S.~Strogatz,
\newblock {\em Nonlinear Dynamics and Chaos: With Applications to Physics,
  Biology, Chemistry, and Engineering} (CRC Press, 2018).

\bibitem{Coppola2021}
S.~Coppola and V.~Kantsler,
\newblock {\em Green algae scatter off sharp viscosity gradients}, Sci. Rep.
  {\bf 11}, 399 (2021).

\bibitem{stehnach_viscophobic_2021}
M.~R. Stehnach, N.~Waisbord, D.~M. Walkama, and J.~S. Guasto,
\newblock {\em Viscophobic turning dictates microalgae transport in viscosity
  gradients}, Nat. Phys. {\bf 17}, 926 (2021).

\bibitem{murata_photoinhibition_2007}
N.~Murata, S.~Takahashi, Y.~Nishiyama, and S.~I. Allakhverdiev,
\newblock {\em Photoinhibition of photosystem {II} under environmental stress},
  Biochim. Biophys. Acta - Bioenerg. {\bf 1767}, 414 (2007).

\bibitem{takahashi_how_2008}
S.~Takahashi and N.~Murata,
\newblock {\em How do environmental stresses accelerate photoinhibition?},
  Trends in Plant Science {\bf 13}, 178 (2008).

\end{thebibliography}


\clearpage

\begin{thebibliography}{20}
\expandafter\ifx\csname natexlab\endcsname\relax\def\natexlab#1{#1}\fi
\expandafter\ifx\csname bibnamefont\endcsname\relax
  \def\bibnamefont#1{#1}\fi
\expandafter\ifx\csname bibfnamefont\endcsname\relax
  \def\bibfnamefont#1{#1}\fi
\expandafter\ifx\csname citenamefont\endcsname\relax
  \def\citenamefont#1{#1}\fi
\expandafter\ifx\csname url\endcsname\relax
  \def\url#1{\texttt{#1}}\fi
\expandafter\ifx\csname urlprefix\endcsname\relax\def\urlprefix{URL }\fi
\providecommand{\bibinfo}[2]{#2}
\providecommand{\eprint}[2][]{\url{#2}}

  
\bibitem{polin2009}
M. Polin, I. Tuval, K. Drescher, J. P. Gollub, and R. E. Goldstein, {\it Chlamydomonas Swims with Two ?Gears? in a Eukaryotic Version of Run-and-Tumble Locomotion}, Science 325, 487 (2009).

\bibitem{de_mooij_impact_2016}
T. De Mooij, G. De Vries, C. Latsos, R. H. Wijffels, and M. Janssen, {\it Impact of light color on photobioreactor productivity}, Algal Research 15, 32 (2016).

\bibitem{bright_two-dimensional_1987}
D. S. Bright and E. B. Steel, {\it Two-dimensional top hat filter for extracting spots and spheres from digital images}, Journal of Microscopy 146, 191 (1987).

\bibitem{Vincent1991}
L. Vincent and P. Soille, {\it Watersheds in digital spaces: an efficient algorithm based on immersion simulations}, IEEE Transactions on Pattern Analysis and Machine Intelligence 13, 583 (1991).

\bibitem{giometto_generalized_2015}
A. Giometto, and F. Altermatt, and A. Maritan, and R. Stocker, and A. Rinaldo, {\it Generalized receptor law governs phototaxis in the phytoplankton Euglena gracilis}, Proc. Natl. Acad. Sci. U.S.A. 112, 7045 (2015).

\bibitem{arrieta2017}
J. Arrieta, and A. Barreira, and M. Chioccioli, and M. Polin, and I. Tuval, {\it Phototaxis beyond turning: persistent accumulation and response acclimation of the microalga Chlamydomonas reinhardtii}, Sci. Rep. 7, 3447 (2017).

\bibitem{ramamonjy2022}
A. Ramamonjy, and J. Dervaux, and P. Brunet, {\it Nonlinear Phototaxis and Instabilities in Suspensions of Light-Seeking Algae}, Phys. Rev. Lett. 128, 258101 (2022).

\bibitem{Coppola2021}
S. Coppola and V. Kantsler, {\it Green algae scatter off sharp viscosity gradients,} Scientific Reports 11, 399 (2021).

\bibitem{stehnach_viscophobic_2021}
M. R. Stehnach, N. Waisbord, D. M. Walkama, and J. S. Guasto, {\it Viscophobic turning dictates microalgae transport in viscosity gradients}, Nat. Phys. 17, 926 (2021).

\bibitem{anguige_one-dimensional_2009}
K. Anguige and C. Schmeiser, {\it A one-dimensional model of cell diffusion and aggregation, incorporating volume filling and cell-to-cell adhesion}, J. Math. Biol. 58, 395 (2009).

\bibitem{baker2008}
N. R. Baker, {\it Chlorophyll Fluorescence: A Probe of Photosynthesis In Vivo}, Annu. Rev. Plant Biol. 59, 89 (2008).

\bibitem{armstrong_continuum_2006-1}
N. J. Armstrong, and K. J. Painter, and J. A. Sherratt, {\it A continuum approach to modelling cell?cell adhesion}, Journal of Theoretical Biology 243, 98 (2006).

\bibitem{murata_photoinhibition_2007}
N. Murata, and S. Takahashi, and Y. Nishiyama, and S. I. Allakhverdiev, {\it Photoinhibition of photosystem II under environmental stress}, Biochim. Biophys. Acta - Bioenerg. 1767, 414 (2007).

\bibitem{leroyer_driftdiffusion_2010}
Y. Leroyer and A. W\"uger, {\it Drift?diffusion kinetics of a confined colloid}, J. Phys.: Condens. Matter 22, 195104 (2010).

\bibitem{strogatz2018nonlinear}
S. Strogatz, {\it Nonlinear Dynamics and Chaos: With Applications to Physics, Biology, Chemistry, and Engineering} (CRC Press, 2018).

\bibitem{hartmann1992}
H. Hartz, C. Nonneng\"asser, and P. Hegemann, {\it The photoreceptor current of the green alga Chlamydomonas}, Phil. Trans. R. Soc. Lond. B 338, 39 (1992).

\bibitem{takahashi_how_2008}
S. Takahashi and N. Murata, {\it How do environmental stresses accelerate photoinhibition?}, Trends in Plant Science 13, 178 (2008).

\bibitem{rorsman2018}
N. J. G. Rorsman, C. M. Ta, H. Garnett, P. Swietach, and P. Tammaro, {\it Defining the ionic mechanisms of optogenetic control of vascular tone by channelrhodopsin-2: Channelrhodopsin-2 in vascular optogenetics}, British Journal of Pharmacology 175, 2028 (2018).

\bibitem{wang2007}
H. Wang, J. Peca, M. Matsuzaki, K. Matsuzaki, J. Noguchi, L. Qiu, D. Wang, F. Zhang, E. Boyden, K. Deisseroth, H. Kasai, W. C. Hall, G. Feng, and G. J. Augustine, {\it High-speed mapping of synaptic connectivity using photostimulation in Channelrhodopsin-2 transgenic mice}, Proc. Natl. Acad. Sci. U.S.A. 104, 8143 (2007).

\bibitem{Lahlou2023}
A. Lahlou, H. S. Tehrani, I. Coghill, Y. Shpinov, M. Mandal, M.-A. Plamont, I. Aujard, Y. Niu, L. Nedbal, D. Lazár, P. Mahou, W. Supatto, E. Beaurepaire, I. Eisenmann, N. Desprat, V. Croquette, R. Jeanneret, T. Le Saux, and L. Jullien, {\it Fluorescence to measure light intensity}, Nat. Methods 20, 1930 (2023).


\end{thebibliography}
\bibliographystyle{myh-physrev3}

\begin{acknowledgments}
R.J., N.D. and I.E. warmly thank Julien Tailleur for the insightful discussion on active phase separation, for suggesting to change the height of the LED ring and for the critical reading of the manuscript. I.E. thanks Michael Shelley for interesting discussions on the mapping of the model with the porous medium equation. R.J. thanks Adrien Izzet and Annie Colin for providing use of their rheometer. R.J., N.D. and I.E. also thank Julie Plastino for proofreading the English language of the manuscript. This work was supported through the Junior Research Chair Programme (ANR-10-LABX-0010/ANR-10-IDEX-0001-02 PSL; R.J.) and an ED-PIF doctoral fellowship (I.E.).
\end{acknowledgments}

\section*{Author contributions}
R.J. and N.D. devised the research. I.E. performed all the experiments and analyzed the data. A.L'H. performed preliminary experiments and helped in establishing the image analysis pipeline. R.J. developed the theoretical model. S.B. and B.B. provided the tools for the physiological measurements and helped in performing them. A.L. and T.L.S. helped in the measurements of the cell density in the dense phase and in the measurements of the intensity fields. I.E., N.D. and R.J. wrote the manuscript. All authors read and discussed the manuscript. 

\section*{Methods}

\subsection{Cell strain and culture} 
Wild-type CC125 strain and eye-deficient mutant CC1101 were obtained from the Chlamydomonas Resource Center. Cells were grown in Tris-Acetate-Phosphate medium (TAP) \cite{Harris2009} at $\sim 70 {\rm \mu mol/m^2/s}$ and $25^{\circ} {\rm C}$  on an orbital shaker at 140rpm. Cells were synchronized by growing them in day/night illumination cycles (16h/8h). Cells were harvested in the exponential phase (concentration between $\sim 0.4$ and $\sim 5\times10^6 {\rm cells/mL}$) and placed for 20 minutes in the dark with aeration (in the open Petri dish used for the experiments, Falcon 353001) before conducting the experiments. 

\subsection{Setup and visualisation}
Light sources are flexible white LED stripes (16 individual LEDs spaced 16.7mm apart, RS Pro 153-3644) glued on the inner edge of a 90mm diameter Petri dish. We visualized the small Petri dish containing the cell suspension using a red collimated light (Thorlabs M730L5 730nm) that doesn't interfere with phototaxis and a CMOS camera (IDS, UI-3370CP-M-GL) mounted with an objective Tamron (ref .002649). See SI sections S7 and Fig. S8 for a characterization of the light sources. 

\subsection{Measurement of the chlorophyll fluorescence ratio}
High light treatment was achieved by 30min illumination in the setup (Fig. \ref{fig:1}A), with a 460nm blue LED stripe at $78{\rm W/m^2}=303 {\rm \mu mol/m^2/s}$ (JKL Components  - ZFS-85000HD-B). Cells quickly phase-separated into a dense phase, which remained stable during the whole HL treatment. A shaken suspension (100rpm) that cannot phase separate was submitted to the same light conditions to compare the evolution of the fluorescence ratio $F_v/F_m$ which was followed using a fluorescence CCD camera recorder (SpeedZen, JBeamBio, France) present in the laboratory of S.B. and B.B. in Institut de Biologie Physico-Chimique \cite{berne_peculiar_2018}. A dark condition was also set as a control.
During low light relaxation, cells were stirred at 140rpm under low intensity white illumination ($\sim 8 {\rm \mu mol/m^2/s}$). The evolution of the ratio $F_v/F_m$ was followed by sampling directly from the Petri dishes along the course of the experiment. For assessing cells in the dense phase $5 {\rm \mu L}$ was pipetted directly in the branches or the drop and diluted into $65 {\rm \mu L}$ of centrifuged medium culture, while $70 {\rm \mu L}$ was directly pipetted in the dilute conditions.

\subsection{Microscope movies}
Microscopy movies were done on inverted Olympus microscopes (IX73 or IX83) with a camera IDS UI-3000SE-M-GL or Hamamatsu Orca Fusion BT C15440-20UP and a $\times 4$ magnification objective. A 780nm long-pass filter (Thorlabs FGL780S) was placed on the light path to prevent phototactic stimulation of the cells by the visualization light. 

\subsection{Viscosity experiments}
The viscosity of the medium was adjusted by adding Methylcellulose to TAP. Methylcellulose M0387 was purchased from Sigma-Aldrich (1500cP). See SI section S8 and Fig. S9 for details.

\newpage

\section*{Supplementary materials}

\renewcommand{\thetable}{S\arabic{table}}

\setcounter{table}{0}
\renewcommand{\arraystretch}{1.5}

\renewcommand{\thesection}{S\arabic{section}}

\setcounter{section}{0}

\section{Measurement of $k_s$}

In order to measure the branch wavenumber at the onset of instability, we first binarized the images by performing two kinds of thresholding. The first one was an Otsu thresholding using the same threshold for the whole image (threshold computed automatically from the histogram) which allows to  extract  the central blob and large branches. However this type of thresholding misses small branches (the ones brighter than the central blob). Therefore in order to also extract small branches we also applied an adaptive thresholding, where the threshold is computed automatically on small windows of width $31$pixels ($\sim 500 {\rm \mu m}$) browsing the image. The final binary image is then the sum of the two thresholdings (Fig. \ref{FigS1}A-B). 

The masks were then skeletonized using the Lee method implemented in Python scikit library. We then counted the number of branches along concentric circles and divided by the perimeter to obtain the branch density $1/\lambda_s$ and the wavenumber $k_s=2\pi/\lambda_s$ (following notations of the manuscript). We measured and averaged the wavenumber over a $7.5$s duration (pink shading in Fig. \ref{FigS1}C), $15$s after the peak in Fig. S1C (corresponding to the appearance of a not fully-defined pattern, Fig. \ref{FigS1}A(i)). Because the branch density is more or less invariant along radii far enough from the center and the edge of the pattern (Fig. \ref{FigS1}D), we averaged the wavenumber over a ring between 50\% and 90\% of the total radial length of the pattern (blue shaded area in Fig. \ref{FigS1}B and \ref{FigS1}D).


\section{Diffusion coefficient measurement}

In order to measure the run-and-tumble diffusion coefficient of the algae in the dark $D_0$, we made use of two linear LED arrays placed on each side of a rectangular chamber containing the algal suspension in order to first concentrate the cells into a band (Fig. \ref{FigS2}A) and record its spreading dynamics as we turn off the side lights. From the extraction of the vertically averaged density profiles (Fig. \ref{FigS2}B) in the region delimited by the pink rectangle in Fig. S2A (after rotating the frame to align the band with the vertical), we could compute both the density gradients $\partial \rho/\partial x $ and the flux $\Phi(x)=-\int^x \partial \rho/\partial t dx' $. For a diffusive process these two quantities are linearly related via the diffusion coefficient $D_0$, which is what we obtained in the 3 replicas we have performed (Fig. S2C). The density gradient and flux have been computed only in the region depicted by the green rectangle in Fig. \ref{FigS2}B, where the cell concentration was sufficiently small to consider that cells don't interact with each other. Fitting the curves in Fig. \ref{FigS2}C by a linear relation, we finally obtained $D_0=1.7\pm 0.4 \times 10^4 {\rm \mu m^2/s}$ (error = standard deviation on the 3 replicas), consistent with previous measurements \cite{polin2009}.



\section{Estimation of the density of the dense phase}

In order to estimate the density of the algae in the branches and in the final drop, we first needed to characterize their dimensions, i.e the thickness and the area of the pattern. 

To estimate the height of the dense phase (in the final drop or in branches) we imaged the Chlorophyll fluorescence of the algae at different focal positions and quantified the number of algae in focus at each position. The pattern-inducing set-up was placed on an epifluorescence microscope (Olympus IX83) equipped with a LED source (pE-$300^{\rm white}$\textsuperscript{\textregistered}, CoolLED), a filter cube (585 nm excitation filter - FF01-585/29-25 Semrock;  605 nm dichroic mirror - DMLP605R Thorlabs; 775nm emission filter - FF01-775/140-25 Semrock) and a 10x objective (Olympus, UPlanFLN, N.A. 0.3). Fluorescence micrographs of the drop/branches were recorded at different height positions by moving the objective by steps $dz=5 {\rm \mu m}$ which correspond to shifts of focus $dz_f = n_{\rm water}/n_{\rm air}dz=1.33dz$ between each step (taking into account the refraction from water). The yellow-orange excitation wavelength we used corresponds to a range where light absorption from algae is minimized \cite{de_mooij_impact_2016} which allows to excite fluorescence over the full depth of the drop. The wavelength range used to collect fluorescence emission (above 700 nm) minimized reabsorption of the emitted light by the algae and allowed to collect images over the whole depth of the pattern. Moreover, the pattern was here induced using a ring of blue LEDs (instead of white) so that it is filtered by the emission filter and doesn't disturb the imaging. Since the phototactic response of the cells is still lightly triggered at the excitation wavelength (585nm), the excitation light was only turned on during the short exposure time of the images (otherwise the dense phase would be disturbed). 

For each micrograph (Fig. \ref{FigS4}A) we then counted the number of algae in the focal plane. Adapting from Bright \textit{et al} \cite{bright_two-dimensional_1987}, for each depth $z$, we first convoluted the image $Im(z)$ with a disk kernel of radius $r$, $K(r)$, corresponding to the dimension of an alga in the image (Fig. \ref{FigS4}B). The convoluted image was then thresholded to retrieve a map of the positions of the algae in focus on the image and a watershed segmentation \cite{Vincent1991} was next used to finally count the number of algae focused in each image (Fig. \ref{FigS4}C). We used a kernel of radius $r=3.25 {\rm \mu m}~(5 {\rm pixels})$ to yield the distributions in Fig. \ref{FigS4}D. Here are shown the vertical cell distributions in different regions of two different drops and one branch. Green and red curves correspond to two different values for the thresholding, showing no significant differences. The distributions were fitted by the following analytical function: 
\begin{align}
\label{fit_profile}
f(z)=\frac{N_{\rm max}}{2}\Bigg[\erf\bigg(\frac{z-z_1}{w}\bigg)+\erf\bigg(\frac{z_2-z}{w}\bigg)\Bigg],
\end{align}
which describes them very well (full red and dashed green lines in Fig. \ref{FigS4}D). We used this function that represents a square density profile (between $z_1$ and $z_2)$ convoluted with the axial PSF of the imaging system (of depth $w$).
 From these fits, we extract the average thickness $h$ of the final drop and of the branches (Fig. \ref{FigS4}E - circles): $h_{\rm drop}=50\pm5 {\rm \mu m}$ and $h_{\rm branch}=58\pm 5 {\rm \mu m}$. The value of $w\approx 15-20 {\rm \mu m}$ obtained in the different experiments (Fig. \ref{FigS4}E - diamonds) is consistent with our imaging system and therefore supports our interpretation on the shape of the measured profiles. The branches are seen to be slightly larger than the final drop probably because they are not at steady-state, with many cells still arriving from the top liquid layers (as opposed to the drop).

Once we know the thickness of the dense phase, we can estimate its cell density by measuring the area of the final drop for different initial cell concentrations $\rho_0$ and assuming that all the cells have joined the dense phase. 
We have done this for 3 initial cell numbers $N=\rho_0V_{\rm susp}$ ($V_{\rm susp}=2$mL), which leads to a linear dependency between the final drop area (averaged over 50 frames, i.e. 25s) and the initial cell number Fig. \ref{FigS4}F (demonstrating that the thickness is always the same). The slope of this linear relation is $s=6.75\times 10^{-8} {\rm cm^2}$, which means that within an area of the drop $a$ there are on average $N=a/s$ cells. From this, we extract the concentration of the dense drop $\rho_{\rm drop}=N/ah_{\rm drop}=1/sh_{\rm drop}\approx 3.0\pm0.3 \times 10^9 {\rm cells/mL}$. Taking the average radius of the micro-algae to be $3.5 {\rm \mu m}$ leads to a volume fraction $\phi\approx 53\pm 5\%$ for the dense phase. 

%

\section{Effect of LED ring geometry on branches positions}

In our setup the 16 white LEDs on the ring (Fig. 1A main text) are separated by a distance of 16.7mm, leading to an angle between consecutive lights $\psi\approx0.33 {\rm rad}\approx 19^{\circ}$. It is therefore natural to question whether branches appear because of an instability as described in the main text or because the light field produced by the discrete light sources is not homogeneous enough. The question is then whether the locations of the branches correlate with the discrete positions of the light sources. By superimposing a snapshot of the pattern (obtained at the same time after turning on the LEDs) for 3 replicas with the same algal culture and the same LED configuration, we can observe that the pattern is never the same (Fig. \ref{FigS5}A): the branch density doesn't change in each replica but individual branches do not appear at the same place. Moreover, the fact that the branch density evolves with our control parameters (Fig. 3A-C main text) further demonstrates that there is no correlation between the LED positions and the locations of the branches in this setup. Those imply that the light field produced by our setup is homogeneous enough along the polar angle. This was confirmed by performing fluorescent measurements to quantify the total light field within the Petri dish containing the cells (section S14 and Fig. \ref{FigS12}).

On the contrary, if we reduce the number of LEDs on the ring by sticking an opaque tape on some of the light sources, then branches only localize in the angular regions facing the taped LEDs, Fig. \ref{FigS5}B. In that situation branches do not appear because of the instability described in the main text, but simply because the light field is not homogeneous enough. Here only 4 equally spaced LEDs were shining on the algal suspension.

%

\section{Measurement of absorption coefficient/characteristic length scale of light propagation}

In order to estimate the characteristic length scale of light propagation in the dense phase, we first measured the absorption coefficient per unit length and cell density of the cell suspension (in the dilute limit) by classical OD measurements (using a Biospectrometer\textsuperscript{\textregistered} Basic from Eppendorf). We measured the absorbance $A=\log(I_0/I_T)$ at 460nm (corresponding to the peak sensitivity of the channelrhodopsin molecules of the eyespot) in a cuvette of length $L_{\rm cuv}=1 $cm at different cell concentrations $\rho_0$ between $\sim 0.3$ and $6\times 10^6 {\rm cells/mL}$ and obtained a linear relationship consistent with a simple Beer-Lambert law for light absorption by the cells (Fig. \ref{FigS6}):

\begin{align}
\label{Beer-lambertl}
A=\log(I_0/I_T)= \kappa_{\rho}\rho_0 L_{\rm cuv}
\end{align}
with $\kappa_{\rho}$ the absorption coefficient per unit length and cell density of the cell suspension. Fitting the experimental absorbance curve returned $\kappa_{\rho}=3\times10^{-7} {\rm cm^2}$. 

From this measurement we extrapolate to the high cell density regime in order to extract a characteristic length scale for light propagation in the dense phase: 
\begin{align}
\label{Beer-lambertl}
L_{\rm abs}=1/\kappa_{\rho}\rho_{\rm drop}\approx10 {\rm \mu m}.
\end{align}
This length corresponds to the width of $\sim$ two cells, showing that a single cell absorbs already a significant quantity of light. 

%
%
%

\section{Theoretical model}

We modelled the density instability presented above with an unsteady drift-diffusion equation for the cell density field as commonly done for phototaxis \cite{giometto_generalized_2015,arrieta2017,ramamonjy2022}. 
Although the light incidence angle (Fig. \ref{FigS3}) imposed a vertical component on top of the radial migration, branches appeared because of a component in the azimuthal direction due to shading effects by surrounding density fluctuations, as a consequence of the cell's eyespot that laterally scans the environment as the cell spins along its swimming axis (on average along the radial direction). As justified below, we neglected the radial and vertical migrations over the time-scale of branch development (typically 1 min).

\subsection{Justification of the 1D approach}


Before the onset of pattern formation, cell density is homogeneous. Right after switching the stimulating light on, cells homogeneously migrate towards the center. As long as their velocity is constant and homogeneous, the cell density at a distance $r$ increases linearly with time as a natural consequence of the cylindrical geometry. The relative increase in cell density can be estimated as $\delta \rho/\rho_0=vt/r$ with $v\approx 100 {\rm \mu m/s}$ the phototactic speed towards the center, leading to a typical increase of only $\times 1.6$ after $t=60$s at $r=1$cm ($\sim$ half a Petri dish radius). So over the duration of the formation of the pattern ($\sim 60$s) the radial migration doesn't modify significantly the cell density field. 

Secondly, we can estimate what would be the accumulation of cells at the bottom of the dish from their vertical migration if there was no phase separation arising (i.e. no interaction between the cells).  We can consider a simple 1D drift-diffusion process with constant drift speed $|u_z|=|-v\sin(\gamma)|\approx 16.5 {\rm \mu m/s}$ ($\gamma$ is the typical light-incidence angle, Fig. \ref{FigS3}) and constant diffusion $D_0\approx1.7\times10^{4}{\rm \mu m^2/s}$ as introduced earlier. Following \cite{leroyer_driftdiffusion_2010} we can analytically solve for the unsteady drift-diffusion equation $\partial \rho/\partial t=D_0\partial^2 \rho/\partial z^2 - u_z \partial \rho/\partial z$ (with the z-coordinate against gravity and $u_z<0$).  The steady-state is an exponential profile with characteristic length scale $l=D_0/|u_z|\approx 1$mm reached after a characteristic time-scale $s=l/|u_z|\approx 62$s, leading to a peak density at the bottom of the dish $\rho_{\rm bottom}\approx 2.3\rho_0$ only. So again the vertical migration is expected to have only a minor effect on the cell density field over the time-scale of pattern formation.

\subsection{1D drift-diffusion model}


We consider a 1D model based on a drift-diffusion equation with the drift velocity coupled to the density field, aiming to represent the coupling between density and light fields through phototaxis and light absorption by the cells in the real system: 
\begin{align}
\label{Drift-Diffusion}
&\frac{\partial \rho}{\partial t}=D\frac{\partial^2 \rho}{\partial x^2}- \frac{\partial \rho v_{\rm photo}}{\partial x}
\end{align}
with $D$ the diffusivity. We write the drift velocity $v_{\rm photo}$ as: 
\begin{align}
\label{V_photo}
v_{\rm photo}(x,t)&=\alpha q\\
q&=\int_x^{x+\Delta x} \rho(x',t)dx' - \int_{x-\Delta x}^{x} \rho(x',t)dx'
\end{align}
where $x$ corresponds to the curvilinear axis along the azimuthal direction (at any distance $r$ far enough from the edge and the center of the dish, see definition in Fig. 1B(i)). Here the population density $\rho(x,t)$ evolves through a drift term $v_{\rm photo}$ that compares the left and right asymmetry in cell density over a finite distance $\Delta x$. By doing so, we avoided to exactly model light propagation in the suspension
and only considered the left-right asymmetry in eyespot stimulation that bias the motion of individual cells towards the direction of largest density (i.e. of lowest intensity).
In addition, these modulations in local cell density can only be detected within a finite distance $\Delta x$. The parameter $\alpha$ models the sensitivity at which cells detect these left-right density modulations and is expected to depend on the light intensity $I_{\rm LED}$, the absorption coefficient of the cells $\kappa_{\rho}$ or the cell strain. We also added a diffusion $D$ to model noise in the ability of cells to follow specific directions. In the following $D$ will be taken as the run-and-tumble diffusivity of the cells in the dark (a discussion on this choice is proposed in section S11). This diffusion has a stabilizing effect that competes with the destabilizing drift term. Interestingly, as shown in SI section S6.5, this model can be mapped onto the porous medium equation when Taylor expanding for small interaction length $\Delta x$, which is also found when e.g. coarse-graining discrete models of diffusing adhesive particles  \cite{anguige_one-dimensional_2009}. Finally, we can note that our full theoretical description is very similar to the one adopted in \cite{armstrong_continuum_2006-1} as a continuum description of aggregation from cell motion induced by cell-cell adhesion forces.

We first normalize the equations in order to work with non-dimensional quantities.  We write $\tilde{x}=x/L$, $\tilde{t}=D t/L^2$ and $\tilde{\rho}=\rho/\rho_0$, with $L$ the total length of the system (a perimeter at any radial distance $r$ far from the edge and the center of the dish). 
This transformation leads to: 
\begin{align}
\label{Drift-Diffusion_norm}
\frac{\partial \tilde{\rho}}{\partial \tilde{t}}&=\frac{\partial^2 \tilde{\rho}}{\partial \tilde{x}^2}- \tilde{\beta}\frac{\partial \tilde{\rho} \tilde{q}}{\partial \tilde{x}}\\
&=\frac{\partial^2 \tilde{\rho}}{\partial \tilde{x}^2}- \tilde{\beta}\Big[\tilde{q}\frac{\partial \tilde{\rho}}{\partial \tilde{x}}+\tilde{\rho}\frac{\partial \tilde{q}}{\partial \tilde{x}}\Big]  \\
\tilde{q}&=\int_{\tilde{x}}^{\tilde{x}+\Delta\tilde{x}} \tilde{\rho}(\tilde{x}',\tilde{t})d\tilde{x}' - \int_{\tilde{x}-\Delta\tilde{x}}^{\tilde{x}} \tilde{\rho}(\tilde{x}',\tilde{t})d\tilde{x}'=\frac{q}{\rho_0 L}
\end{align}
with $\tilde{\beta}=\alpha\rho_0 L^2/D$ a non-dimensional parameter that controls the emergence of instabilities as we will see in the next section. 


\subsection{Linear stability analysis}


We then performed a linear stability analysis of this set of equations. Injecting solutions of the type $\tilde{\rho}(\tilde{x},\tilde{t})=1+\epsilon e^{\tilde{\sigma}\tilde{t}}e^{-i\tilde{k}\tilde{x}}$ ($\epsilon\ll 1$) into Eq. \ref{Drift-Diffusion_norm} and keeping only terms of order $\epsilon$ \cite{strogatz2018nonlinear}
\begin{align}
\label{terms_eps}
\frac{\partial \tilde{\rho}}{\partial t}&= \epsilon\tilde{\sigma} e^{\tilde{\sigma} \tilde{t}} e^{-i\tilde{k}\tilde{x}}\\
\frac{\partial^2 \tilde{\rho}}{\partial \tilde{x}^2}&=-\epsilon\tilde{k}^2 e^{\tilde{\sigma} \tilde{t}}e^{-i\tilde{k}\tilde{x}}\\
\tilde{\rho}\frac{\partial \tilde{q}}{\partial \tilde{x}}&=2\epsilon \big[\cos(\tilde{k}\Delta \tilde{x})-1\big]e^{\tilde{\sigma}\tilde{t}}e^{-i\tilde{k}\tilde{x}}+O(\epsilon^2)\\
\tilde{q}\frac{\partial \tilde{\rho}}{\partial \tilde{x}}&=O(\epsilon^2),
\end{align}
the dispersion relation is directly obtained (simplifying all $\epsilon$, $e^{\tilde{\sigma}\tilde{t}}$ and $e^{-i\tilde{k}\tilde{x}}$):

\begin{align}
\label{Disp_rel}
\tilde{\sigma}&=2\tilde{\beta}\big(1-\cos(\tilde{k}\Delta \tilde{x})\big)-\tilde{k}^2,
\end{align}
which expresses the growth rate of the instability $\tilde{\sigma}$ for the spatial mode $\tilde{k}$. Coming back to dimensional variables we get: 

\begin{align}
\label{Disp_rel}
\frac{\sigma}{D\beta}&=2\big(1-\cos(k\Delta x)\big)-\frac{k^2}{\beta}\\
&=f(k)-g(k)
\end{align}
where $\beta=\alpha\rho_0/D$ and $f$ and $g$ positive functions of $k$. The system is linearly unstable if and only if there exists a range of wavenumber $k$ for which the growth rate $\sigma$ is positive, i.e. $f(k)>g(k)$. In that case the selected mode $k_s$ corresponds to the mode that maximizes $\sigma$ (i.e. the most unstable). By expanding the function $f$ to second order in $k$ ($f\sim 2 (1-(1-k^2\Delta x^2/2))\sim \Delta x^2 k^2$), we directly see that this is the case when $\Delta x^2>1/\beta$, i.e. $\beta>\beta_c=1/\Delta x^2$.
 
When $\beta \le \beta_c= 1/\Delta x^2$, $f(k)\le g(k)$ for all $k$ (Fig. \ref{FigS6bis}D-cyan and light blue curve), the growth rate $\sigma$ is always negative and the system is linearly stable. This situation was met experimentally when cell concentration or LED intensity is low (Fig. 3A-B, movies S7-8), or when using the eyespot mutant (movie S10), where cells migrated towards the center but didn't develop branches. 

%
%

When $\beta$ increases above $\beta_{c}$, $f(k)>g(k)$ for some range of $k$ (Fig. \ref{FigS6bis}D-dark blue and pink curves) and the system becomes unstable (within the shaded areas of Fig. \ref{FigS6bis}).

%
%
%

\subsection{Mode selection}

When the system is linearly unstable ($\beta>\beta_c$), the mode that is selected corresponds to the one with the highest growth rate $\sigma$, i.e. the most unstable. By taking the derivative of $\sigma$ with $k$, it is straightforward to show that the locations of the extrema of $\sigma$, $k_e$, are given by the relation: 
\begin{align}
\label{Most_unstable}
{\rm sinc}(k_e \Delta x)=\frac{1}{\beta \Delta x^2},
\end{align}
from which we can extract the global maximum:
\begin{align}
\label{Most_unstable2}
k_{s}^{\rm (th)}&=\frac{1}{\Delta x}{\rm sinc}_{[0;\pi]}^{-1}\Big(\frac{1}{\beta \Delta x^2}\Big),
\end{align}
where ${\rm sinc}_{[0;\pi]}^{-1}(y)$ $(y\in [0,1])$ is the inverse function of the sine cardinal ${\rm sinc}(x)$, $x \in [0, \pi]$. The most unstable mode is a monotonically increasing function of $\beta$. When $\beta$ is much larger than $1/\Delta x^2=\beta_c$, $k_s^{\rm (th)}$ tends to a finite value simply given by $k_{s,{\rm max}}^{\rm (th)}=\pi/\Delta x$.


\subsection{Mapping onto the porous medium equation}

Here we show how our drift-diffusion framework can be mapped onto the porous medium equation. 
Starting from the expression of the phototactic drift $ v_{\rm photo} = \alpha[\int_x^{x+\Delta x} \rho(x',t)dx' - \int_{x-\Delta x}^{x} \rho(x',t)dx'] $, we perform a first order Taylor expansion of $\rho$ : $ \rho(x',t) = \rho(x,t) + (x'-x)\frac{\partial{\rho}}{\partial {x'}}(x,t) + o(x-x') $, to get:
\begin{align*}
	\int_x^{x+\Delta x} \rho(x',t)dx' &=  \rho(x,t)\Delta x + \frac{\partial{\rho}}{\partial {x}}(x,t) \frac{\Delta x^2}{2} + o(\Delta x^2) \\
	\int_{x-\Delta x}^{x} \rho(x',t)dx' &=  \rho(x,t)\Delta x - \frac{\partial{\rho}}{\partial {x}}(x,t) \frac{\Delta x^2}{2} + o(\Delta x^2) \\
\end{align*}
and $ v_{\rm photo} = \alpha\frac{\partial{\rho}}{\partial {x}}(x,t) \Delta x^2 + o(\Delta x^2) $.

Injecting this expression in the drift-diffusion equation:
\begin{align*}
	\frac{\partial \rho}{\partial t}&=D\frac{\partial^2 \rho}{\partial x^2}- \frac{\partial \rho v_{\rm photo}}{\partial x} \\
	\frac{\partial \rho}{\partial t}&=\frac{\partial}{\partial x}[D\frac{\partial{\rho}}{\partial {x}} - \alpha\Delta x^2\rho\frac{\partial{\rho}}{\partial {x}}] \\
	\frac{\partial \rho}{\partial t}&=\frac{\partial}{\partial x}[D_{sd}(\rho)\frac{\partial{\rho}}{\partial {x}}] \\
\end{align*}
we get a non-linear diffusion equation with a state-dependent diffusivity $ D_{sd}(\rho) = D-\alpha\Delta x^2\rho $. 

The diffusivity can become negative (and the system is unstable) if $D < \alpha\Delta x^2\rho \Leftrightarrow \frac{\alpha\rho}{D}  > \frac{1}{\Delta x^2}  \Leftrightarrow \beta > \beta_c $. We find the same criterion for system stability as in the linear stability analysis in the original model.
We can also see immediately that the system can be unstable only for negative phototaxis. Indeed for positive phototaxis, $v_{\rm photo +}=-v_{\rm photo}$ and $ D_{sd}(\rho) = D+\alpha\Delta x^2\rho $ remains always positive. 

Moreover, we can see that:
\begin{align*}
	\frac{\partial{D_{sd}}}{\partial t}  &= - \alpha\Delta x^2 \frac{\partial \rho}{\partial t}  \\
	\frac{\partial{D_{sd}}}{\partial x}  &= - \alpha\Delta x^2\frac{\partial \rho}{\partial x}  \\
\end{align*}
and rewrite the previous equation as:
\begin{align*}
 \frac{\partial{D_{sd}}}{\partial t} &= \frac{\partial}{\partial x}[D_{sd}\frac{\partial{D_{sd}}}{\partial {x}}] \\
 \frac{\partial{D_{sd}}}{\partial t} &= \frac{1}{2} \frac{\partial^2 D_{sd}^2}{\partial x^2} \\
\end{align*}
which is the porous medium equation for the state-dependent diffusivity $D_{sd}(\rho)$. 
Taylor-expanding for small interaction lengths $\Delta x$ amounts to assume a purely local density-dependent motility of the cells. By doing so, we retrieve the phase separation at critical densities but this new equation does not lead to finite wavelength selection.

\section{Measurement of LED power}

We describe here how we measured the emitted light power of our white LEDs. For this purpose we used the same LED array as in the instability experiments, but where we blocked all LED except one. As in the experiments, the LED array was controlled by a Thorlabs driver LEDD1B, delivering a voltage $U$ in the range $\sim7-14$V. We then measured the power spectrum of the single LED for all the voltages $U$ used in the instability experiments (LPS 220 spectrofluorometer, Photon Technology International; Software, PTI Felix 4.1.0.3096). An example of measured spectrum (voltage $U=10$V) is shown in Fig. \ref{FigS8}A-black curve. 

Because the power unit of the spectrometer was not calibrated, we then performed intensity measurements with the same LED. We placed a $[430$-$475]$nm band-pass filter (FF01-452/45-25, Semrock) in front of the LED and used a S130C detector together with a PM100D power-meter from Thorlabs to measure the light intensity in this wavelength range at different distances $r$ from the light source. Doing this for two different voltages, $U_{\rm low}=10$V and $U_{\rm high}=14.6$V, returns experimental data consistent with a $r^{-2}$-dependency of the light intensity (Fig. \ref{FigS8}B - black curves). By fitting these curve ($I_{430/475}(r)=P_{430/475}/r^2$) we extract the absolute light power $P_{430/475}$ emitted by the single LED in the range $[430$-$475]$nm. 

We then simply computed the area $A_{430/475}$ of the measured power spectrum in the range $[430$-$475]$nm (black shading in Fig. \ref{FigS8}A) at both voltages $U_{\rm low}$ and $U_{\rm high}$, which gives a linear (i.e. not affine) relationship with $P_{430/475}$. This linear relationship was then used to compute the total light power $P_{\rm LED}$ emitted by a single LED at any voltage $U$, by simply computing the area of the power spectrum over the full range of wavelength. We then define the average intensity per LED within the Petri dish $I_{\rm LED}$ as $I_{\rm LED}=1/(r_{\rm max}-r_{\rm min})\int_{r_{\rm min}}^{r_{\rm max}} P_{\rm LED}/r^2 dr=P_{\rm LED}/(r_{\rm min} r_{\rm max})$,
where $r_{\rm min}=3.25 {\rm cm}$ is the distance from a LED to the edge of the dish and $r_{\rm max}=5.0 {\rm cm}$ the distance to the center of the dish.

For the blue LEDs used in the experiments to quantify photo-protection, we extracted the light power at the voltage used in these experiments by measuring the distance dependency of the light intensity of a single LED at $460$nm (peak in LED spectrum, Fig. \ref{FigS8}A - blue curve) and fitting by $I(r)=P_{\rm blue}/r^2$ (Fig. \ref{FigS8}B - blue curve). In the manuscript, we report the intensity value $I_{\rm tot}=78{\rm W/m^2}=303 {\rm \mu mol/m^2/s}$ which corresponds to the total average light intensity within the Petri dish, estimated by considering all 32 LED of the array. 

We converted the ${\rm W/m^2}$ unit into ${\rm \mu mol/m^2/s}$ through the following formula $I_{\rm \mu E}=I_{W/m^2} \lambda/(10^{-6}hc N_A)$ for the blue monochromatic light, or averaging over the power spectrum for the white light.  $h$ is the Planck constant, $c$ the vacuum light speed and $N_A$ the Avogadro number.


\section{Viscosity experiments}

In order to increase the viscosity of the medium, we simply dissolved some Methylcellulose (MC, purchased from Sigma-Aldrich M0387, 1500cP) in a stock of TAP medium at relatively large concentration to make a mother solution. We then diluted our cell suspension, with some amount of MC-TAP solution and some amount of fresh TAP medium in order to make cell suspensions with the same cell concentration but different MC concentration and therefore different medium viscosity. 

The viscosity of different MC solutions was measured with a rheometer (Discovery HR-1, TA instruments) in a cone-plate configuration by making a slow ramp in shear rate and measuring the torque. Viscosities as a function of shear rates for all solutions are shown in Fig. \ref{FigS9}A. Our MC solutions clearly display some shear-thinning behavior at low shear rate, until reaching a well-defined plateau above $\sim 50 {\rm s^{-1}}$. Averaging the viscosity over the plateau returns the data shown Fig. \ref{FigS9}B for the viscosity as a function of MC concentration (expressed as a percentage of the mother MC solution added). The evolution of the viscosity is compatible with an exponential dependency of the MC concentration, as previously reported. We used the fitted analytical function $\eta([MC])=\eta_0\exp([MC]/b)$ ($\eta_0=0.001$Pa.s fixed parameter and $b=33.0\pm0.6$\% fitting parameter) to then estimate the viscosity of all the solutions used in the instability experiment. 

The range of shear rates over which our MC solutions exhibit a Newtonian behavior is relevant for the beating of the flagella ($\sim 60$Hz). If the force developed by the pair of beating flagella doesn't change with increasing viscosity, the speed of the cells is expected to be inversely proportional to the drag coefficient and therefore to the viscosity. This was indeed measured to be the case in \cite{Coppola2021,stehnach_viscophobic_2021} for $\eta \lesssim 5 \eta_0$.

\section{Qualitative description of the evolution of the experimental patterns}

When increasing cell density (at constant LED intensity $I_{\rm LED}=6.2 {\rm W/m^2}$), branches got closer to each other and became wider (Fig. 3A and \ref{FigS10}A, movie S7). Instead, when the light power was increased (at a fixed concentration $\rho_0=1.2\times 10^6 {\rm cells/mL}$) the branches got closer but became thinner (Fig. 3B and \ref{FigS10}B, movie S8). Moreover, below some value of $\rho_0$ ($<3\times10^5$cells/mL) or $I_{\rm LED}$ ($<32 {\rm mW/m^2}$) cells still performed negative phototaxis but did not form any pattern (and $k_s=0$, Fig. 3A-B, movies S7-8), as is the case when using an eyespot deficient mutant at any light intensity (CC1101, movie S10). 
The evolution of $k_s(\rho_0)$ and $k_s(P_{LED})$ corresponds to what can be expected from the instability mechanism proposed in the manuscript: i) as density fluctuations become more prominent at high cell density, the suspension is more prone to destabilization, ii)
at high light intensity, the flux of photons to the eyespot increases and cells' ability to accurately follow specific light directions increases, making them more sensitive to asymmetries in cell density.
When increasing the viscosity of the medium to a few times that of the growth medium (at fixed concentration $\rho_0=1.6\times 10^6 {\rm cells/mL}$ and LED intensity $I_{\rm LED}=6.2 {\rm W/m^2}$), branches also got closer to each other and became thinner (Fig. 3C and \ref{FigS10}C, movie S9). 

Regardless the control parameter that was varied ($\rho$, $I_{\rm LED}$, $\eta$), the wavenumber seemingly saturated at a similar value $k_{s,{\rm max}}\sim 7$-$10 {\rm mm^{-1}}$
(corresponding to a minimal average inter-branch distance $\lambda_{s,{\rm min}}\sim 600$-$900 {\rm \mu m}$). This is consistent with a finite interaction length $\Delta x$: as the pattern is emerging, newly formed filaments are still thin enough to let light go through (as in Fig. 1B(ii)) and are therefore able to interact and merge (as in movie S6), as long as they are within the interaction distance. 
After some time, branches are found, on average, further apart from each other than the interaction distance and branch merging stops, even though branches are still thin enough to be able to interact as is the case with large viscosities or high light intensities at sufficiently low cell concentration, (Fig. 3B-C - Insets). 

At cell densities larger than $\sim 6\times 10^6 {\rm cells/mL}$ (i.e. above the range shown in Fig. 3A), we did not obtain nice and well-defined patterns (movie S11). During the initial radial migration of the cells a denser ring formed at $\sim 1$cm from the center of the dish which we believe was also the consequence of light absorption by the cells. Here, because cell concentration is already quite high initially, cells on the periphery of the Petri dish block light for others, limiting their ability to perform negative phototaxis 
and thus creating an accumulation of cells at a given radial distance. Ortho-radial destabilization still happened simultaneously, but the width of the branches became comparable to the minimal inter-branch distance and the pattern cannot be well defined throughout the migration.

\section{Fitting of the experimental data}

\subsection{Model for the theoretical parameters $D$ and $\alpha$}

In order to compare our experimental measurements of the selected mode $k_s$ with our theoretical drift-diffusion framework (Eq. 3 or \ref{Most_unstable2}) we needed to write models for the dependency of $D$ and $\alpha$ with our control parameters $\rho_0$, $\eta$ and $I_{\rm LED}$.

Firstly, because the swimming speed of the algae $U$ has been shown to be inversely proportional to the viscosity of the medium (up to $\sim 5\times$ the viscosity of water) \cite{Coppola2021,stehnach_viscophobic_2021}, and because the run-and-tumble diffusivity (in the dark) is proportional to $U^2\tau$ with $\tau$ the average run time, we made the hypothesis that $D(\eta)=D_0 \eta_0^2/\eta^2$, with $\eta_0$ the viscosity of the TAP medium (equal to that of water $\eta_0=1 {\rm mPa.s}$) and $D_0=1.7\times 10^4 {\rm \mu m^2/s}$ the diffusion coefficient in TAP. This follows from the hypothesis that $\tau$ doesn't depend on viscosity. Moreover, we ignored any dependency of the cell diffusion coefficient with local cell density, which should arise at large cell concentration (because of collisions between the cells). Such effect should matter once the patterns has formed but not much for its initial formation and the mode selection. 

Secondly, because phototaxis is mainly due to light received by the photo-sensitive channelrhodopsins molecules composing the eyespot, we decided to model the phototactic sensitivity $\alpha$ following the light intensity response of these molecules. In different in vitro and in vivo systems, channelrhodopsin molecules have been robustly shown to respond to light intensity following a Hill function which saturates at high intensity \cite{hartmann1992,rorsman2018,wang2007}. Therefore we write 
$\alpha(I_{\rm LED})=aI_{\rm LED}^p/(K^p+I_{\rm LED}^p)$, where $p$ is the exponent of the Hill function, and $K$ the light intensity at which the response is half the maximum value $a$ (which is the saturating value of the phototactic sensitivity).  In channelrhodopsin molecules the exponent $p$ has always been found to be $\sim 0.7$ \cite{hartmann1992,rorsman2018,wang2007}, while the half-saturation constant $K$ appears to be strongly system-dependent. Regarding the parameter $a$, it should probably depend on cell concentration, absorption coefficient of the cells or strain, but we again ignored those to keep the model as simple as possible and limit the number of fitting parameters. 

From these two models we could then write the parameter $\beta$ of the drift-diffusion framework as a function of our experimental and control parameters: 

\begin{align}
\label{beta_param}
\beta&=\frac{a\rho_0}{D_0}\frac{\eta^2}{\eta_0^2}\frac{I_{\rm LED}^p}{K^p+I_{\rm LED}^p},
\end{align}
leading to:

\begin{align}
\label{ks_exp}
 k_{s}^{\rm (th)}(\rho_0,\eta, I_{LED})&= \frac{1}{\Delta x}{\rm sinc}_{[0;\pi]}^{-1}\Big(\frac{D_0}{a\rho_0\Delta x^2}\frac{\eta_0^2}{\eta^2}\frac{K^p+I_{\rm LED}^p}{I_{\rm LED}^p}\Big)
\end{align}

In the following we use this expression to fit our experimentally measured $k_s$.

\subsection{Extracted fitting parameters}

%

Here we detail the way we have fitted each experimental dataset in the parameter space $(\rho_0, I_{\rm LED}, \eta)$. 

For the experimental dataset "density" (when varying $\rho_0$), we wrote $\kappa=aI_{\rm LED}^p/D_0(K^p+I_{\rm LED}^p)$ to have two independent free fitting parameters $\Delta x$ and $\kappa$: 
\begin{align}
\label{fit_dens}
 k_{s}^{(d)}= \frac{1}{\Delta x}{\rm sinc}_{[0;\pi]}^{-1}\Big(\frac{1}{\rho_0 \kappa \Delta x^2}\Big)
\end{align}

For the experimental dataset "viscosity" (when varying $\eta$), we wrote $\mu=aI_{\rm LED}^p/D_0\eta_0^2(K^p+I_{\rm LED}^p)$ to have two independent free fitting parameters $\Delta x$ and $\mu$: 
\begin{align}
\label{fit_dens}
 k_{s}^{(v)}= \frac{1}{\Delta x}{\rm sinc}_{[0;\pi]}^{-1}\Big(\frac{1}{\eta^2 \mu \Delta x^2}\Big)
\end{align}

For the experimental dataset "intensity" (when varying $I_{\rm LED}$), we had more free parameters entering the analytical expression, $\Delta x$, $\nu=a\rho_0/D_0$ and the Hill coefficients $p$ and $K$:  
\begin{align}
\label{fit_dens}
 k_{s}^{(i)}= \frac{1}{\Delta x}{\rm sinc}_{[0;\pi]}^{-1}\Big(\frac{1}{\nu \Delta x^2}\frac{K^p+I_{\rm LED}^p}{I_{\rm LED}^p}\Big)
\end{align}
which makes the fitting procedure ill conditioned. We therefore decided to fix the exponent $p=0.7$ in agreement with the measurements on channelrhodopsins. We also fixed the parameter $\Delta x$ to the average value obtained in the experiments "density" and "viscosity" which were in relatively good agreement. This makes $\nu$ and $K$ the only two fitting parameters for this dataset. Manually changing the fixed parameter $p$ between $0.2$ and $2$ worsened the fits (Fig. \ref{FigS11}), further suggesting that the value $p\approx0.7$ is very robust.

Our goal was then to quantitatively compare, i) the parameter $\Delta x$ in the "density" and the "viscosity" experiments, and ii) the parameter $a$ that we can extract in each dataset from the parameters $\kappa$, $\mu$ and $\nu$, in order to check the validity of our model. The values obtained are shown in table \ref{tabS1} below (fitting parameters shown in gray shading). 

\begin{table}[h!]
\centering
\begin{tabular}{| S{p{2.4cm}} | S{p{2.1cm}} | S{p{1.5cm}} | S{p{2cm}} | S{p{1.5cm}} | S{p{3cm}} | S{p{2cm}} |}
\hline
Experiment & $\rho_0$ (cells/mL) & $\eta~{\rm (mPa.s)}$ & $I_{\rm LED}~{\rm (W/m^2)}$ & $\Delta x~{\rm (\mu m)}$ & Other fit parameters (SI) & a (SI)\\
\hline
Cell density & / & $1$ & 6.15 & \cellcolor[gray]{0.6}$379 \pm 11$ & \cellcolor[gray]{0.6}$\kappa=2.14\pm 0.12\times10^{-5}$ & $3.82\pm0.25\times 10^{-13}$ \\
\hline
Viscosity & $1.6\times10^6$  & / & 6.15 &  \cellcolor[gray]{0.6}$306\pm11$ &\cellcolor[gray]{0.6}$\mu=2.23\pm0.29\times10^{13}$ & $2.48\pm0.33\times10^{-13}$ \\
\hline
LED intensity & $1.2\times10^6$  & $1$ & / & 342.5  \newline (fixed) & \cellcolor[gray]{0.6}$\nu=2.98\pm0.28\times 10^7$ \newline $K=0.078\pm0.019 $ & $4.22\pm0.39\times10^{-13}$ \\
\hline
\end{tabular}
\caption{Parameters obtained from the fitting of each experimental dataset}
\label{tabS1}
\end{table}

The parameter $\Delta x$ is consistent between the two datasets ($\sim 350 {\rm \mu m}$).
The values obtained for $a$ are also consistent between the experiments, we get on average $a=3.5\pm0.9 \times 10^{-13} {\rm m^3/s}$. Overall this shows that our simple model captures the physics of the destabilization process. We can also comment on the saturation of the wavenumber $k_s$ in the light intensity experiment at a smaller value than in the other datasets: from our model this comes from the saturation of the phototactic sensitivity $\alpha$ as intensity is increased well above the parameter $K=0.078 {\rm W/m^2}$: $k_{s,\rm max}^{\rm (th)}= \frac{1}{\Delta x}{\rm sinc}_{[0;\pi]}^{-1}\Big(\frac{D_0}{a\rho_0\Delta x^2}\Big)$ (and not $\pi/\Delta x$ as in the other dataset).

For plotting the phase diagram of the system (Fig. 3E,F) we have taken the averages of the different fitting parameters.

\section{Discussion on the diffusion coefficient $D$}

The diffusivity in the drift-diffusion equation was taken to be the value of the run-and-tumble diffusion in the dark. This choice is questionable since cell migration towards the center is highly directed, and cell dispersion along the azimuthal direction is certainly smaller. However the exact value of $D_0$ in the expression of the selected wavenumber $k_{s}^{\rm (th)}$ is compensated by the fitted value of the parameters $a$, since it is the ratio $D_0/a$ that appears in all fitting parameters $\kappa$, $\mu$ and $\nu$. What is important for explaining finite wavelength selection is the form of the dispersion relation, not much the exact values of the parameters (as long as they are consistent between all experiments). It would be interesting to try predicting what should be the proper diffusivity to be used here, for instance by performing single cell tracking experiments and/or by properly quantifying the growth rate of the selected modes in the experiments. We believe this is out of scope of the present article and we leave this question mark to future investigations. 

\section{Discussion on the interaction length $\Delta x$ and on the validity of the 1D approach}

As discussed in the article, the extracted value of the interaction length $\Delta x$ appears to be consistent with a prediction based on the geometry of the setup $\Delta x_{\rm geo}=h_{\rm branch}/\tan(\gamma)$, where we consider the length of the shading of a branch of height $h_{\rm branch}$ from the side lights shining at a typical angle $\gamma$ (Fig. \ref{FigS11b}). In Fig. 3F of the article, we show three theoretical curves showing the approximate range of admissible curves from the estimated errors of the different parameters, and considering the angle defined as $\gamma={\rm atan}(H_{\rm LED}/L)$ with $L$ varied between $R_{\rm LED}=4.5 {\rm cm}$ and $R_2=4 {\rm cm}$ (see schematics in Fig. \ref{FigS11b}). The central curve, closer to the experimental points, was obtained from the average values. The experimental parameters were: $\rho_0=2.07\pm0.5 \times 10^6 {\rm cells/mL}$, $h_{\rm branch}=58\pm 5 {\rm \mu m}$, $ L=4.25 \pm 0.25 {\rm cm}$ and $\kappa=2.14\pm0.15 \times 10^{-5} {\rm m}$. 

In Fig. 3F we have also gray shaded the region of large LED height. We believe our 1D approach is not valid anymore in this region. In particular the vertical migration towards the bottom should become important and the effective concentration to be considered for branch formation should not be the initial cell concentration $\rho_0$. Performing experiments at higher LED height $H_{\rm LED}=15.5 $cm (movie S13) and above shows indeed that branches still appear on the periphery while our simple approach predicts no branch formation (i.e. $k_s^{\rm (th)}=0$, Fig. 3F). In these experiments, the local increase of cell concentration at the bottom of the dish prior to branch formation probably shifts the effective concentration $\rho_0$ that should be used in the 1D approach. The fact that in these experiments branches appear only in the periphery but not in the central region is consistent with this interpretation, since vertical cell migration must depend on radial position and should be stronger on the periphery than in the central region. 

Again it would be interesting to go further and explore the phase space of the system when systematically varying all experimental parameters, including $H_{\rm LED}$, and to build a full 3D model that would be valid for all experimental configurations.

\section{Chlorophyll fluorescence measurements}

Once sufficiently large and dense, the branches only retracted through their interface with no net phototactic flux in their bulk (Fig. 2), suggesting that the crowded environment of the dense phase could be beneficial for the algae with regards to the harmful effects of excess light on the photosynthetic machinery, the major of which being the photoinactivation of Photosystem II (PSII). Such harmful effects are routinely quantified via the maximal photochemical efficiency of PSII, or $F_v/F_m$ ratio, through chlorophyll fluorescence measurements.

Indeed, energy absorbed within the light-harvesting complexes can be dissipated into fluorescence, heat or photochemistry, 
so that the fluorescence level allows to measure the global quantum efficiency of photochemistry \cite{baker2008}. 
The ratio $F_v/F_m$ is maximal in healthy Chlamydomonas populations (usually 0.75-0.8), and drops when light absorption exceeds the capacity of photon usage by the photosynthetic machinery (which itself depends on CO2 level, temperature and other factors), reflecting photoinactivation of PSII (photoinhibition) \cite{murata_photoinhibition_2007, takahashi_how_2008}. Treatment in low light favors PSII repair over inactivation, allowing to recover a maximal $F_v/F_m$ over a couple of hours, whereas addition of chloroplast translation inhibitors such as lincomycin prevents such repair.  

In order to quantify photoinhibition in our system, we subjected the cells to thirty minutes of strong blue lights (using a LED array similar to the white one) at total average intensity $I_{\rm tot}= 78{\rm W/m^2}=303 {\rm \mu E/m^2/s}$ (spatial average over the Petri dish from the 32 LED of the array) followed by two hours of darkness to evaluate the relaxation after photoinhibition. Powerful blue LEDs were used to ensure photodamage was building up over relatively short time-scales and to avoid later triggering photo-protective mechanisms (eg. high energy quenching qE).
Cells were either let free to phase separate or prevented from it by gently shaking the system during photo-stimulation (at $100$rpm, to homogenize the suspension). We measured the ratio $F_v/F_m$ (Materials and Methods) by collecting small amounts of cells in the branches or in the central drop (depending on the time point) and compared them to shaken suspensions. The ratio decayed from $\sim 0.8$ to $\sim 0.6$ in the dense phase and to $\sim 0.2$ in the shaken suspension during photo-stimulation, while it relaxed to the initial value during darkness, Fig. 4 solid blue and orange curves respectively. 
As expected, relaxation of $F_v/F_m$ in the dark did not take place at all when lincomycin was added (Fig. 4, dashed blue and orange curves). Overall these results show that photoinhibition strongly occurred in the shaken suspension but much less in the phase-separated state. If indeed light penetrates over long distances in the shaken suspension, affecting many cells, it is strongly blocked over the first tens of microns in the dense phase, so that cells within the bulk are not affected.

\section{Measurements of the total intensity field in the setup}

In order to measure the total intensity field (from the sixteen LEDs) in the Petri dish containing the algae, we made use of a recently developed fluorescence technique \cite{Lahlou2023}. Briefly, this method relies on the time decay of the emitted fluorescence light of the molecule Dronpa2 when shining blue light. The exponential decay of the emitted fluorescence light (between $500$ and $600$nm) is inversely proportional to the local light intensity exciting the molecule. Therefore by recording the emitted fluorescence intensity of a solution containing Dronpa2 molecules and fitting the time decay of the intensity of each pixel of the recorded movies, we can access the intensity field of the excitation light. 

In order to quantify the effect of light absorption by the cells on the total intensity field in the Petri dish, we have performed measurements at different concentrations of micro-algae. Because the characteristic time scales for the decay of the fluorescence of Dronpa2 is of the order of a few seconds, we had to use dead cells in the suspension. Otherwise the rapid motion of the cells would advect the molecules which would start to feel a distribution of light intensity, blurring the overall intensity map we would extract. Cells were killed by adding a small amount of Lugol's iodine ($2 \permil$v/v), and subsequently centrifuged and resuspended in TAP in order to remove the Lugol (which has a brown color that would disturb the measurements). A small volume of concentrated solution of Dronpa2 (130 $\mu$M) was then added to the suspensions to reach a final concentration of $\sim 0.3\mu$M. A volume of $2$mL of this suspension was then loaded into the Petri dish. We subsequently excited the fluorescence of the molecule with the illumination setup of the branch instability experiments, replacing the white LED ring by a blue one (same number of LEDs and same inter-LED spacing). The intensity fields obtained this way correspond to the integration over the thickness of the liquid ($\sim 2$mm). A full 3D modeling of the intensity field without cells can be found in the Supplementary Materials of \cite{Lahlou2023}.

The results are shown in Fig. \ref{FigS12} for different concentrations of cells. Those represent the initial conditions of the intensity field in the branch experiments. We can first notice that all the intensity fields are axisymmetric, with almost no trace of the discrete light sources. Moreover, in absence or at low concentrations of cells the intensity field is actually maximum in the center of the Petri dish. This is rather counter-intuitive since the light intensity for each source decays as $1/r^2$. In fact as shown in \cite{Lahlou2023} this mainly stems from a lensing effect of the water-filled Petri dish that refocuses the diverging lights from the LED. Finally, as cell concentration is further increased above $\sim 2\times 10^6 {\rm cells/mL}$ the intensity field becomes maximum on the edge and minimum at the center of the dish. This is due to light absorption by the cells. 

We can rationalize the cell density at which the switch in intensity field happens. The characteristic length scale for light absorption is given by $L_{\rm abs}(\rho_0)=1/\kappa_{\rho_0}\rho_0$, with $\kappa_{\rho}=3\times 10^{-7} {\rm cm^2}$ (section S5). At $\rho_0=10^6 {\rm cells/mL}$ the length scale is then $L_{\rm abs}\approx 3.3$cm which is larger than the radius of the Petri dish (1.75 cm), while at $\rho_0=2\times 10^6 {\rm cells/mL}$ the length scale is halved $L_{\rm abs}\approx 1.65$cm, of the order of the radius of the Petri dish. Therefore we start observing a large influence of the cell absorption on the total intensity field only for $\rho_0\ge 2\times 10^6 {\rm cells/mL}$.

\section{Captions of the Supplementary Movies}

\begin{itemize}[label={$\bullet$}]

\item Movie S1: Movie of the destabilization of the suspension at $\rho_0=2.8 \times 10^6$cells/mL (corresponding to the time-lapse shown in Fig. 1B). The movie is sped up by $\times10$. 

\item Movie S2: Movie of the destabilization seen under a microscope at $\times4$ magnification. The movie is sped up by $\times 4$. 

\item Movie S3: Movie showing the effect of switching off the lights after forming the pattern. The sharp pattern immediately disappears from the run-and-tumble diffusion of the cells in the dark. The movie is sped up by $\times6$.

\item Movie S4: Movie showing the slow retraction of a large branch ($w>w_c$) as seen under a microscope at $\times 4$ magnification. The movie is displayed at real speed. 

\item Movie S5: Movie showing the faster retraction of a thin branch ($w<w_c$) as seen under a microscope at $\times 4$ magnification. The movie is displayed at real speed. 

\item Movie S6: Movie showing that individual branches can get attracted to neighboring ones, because of shading effects. 
The movie was obtained from visualization under a microscope at $\times 4$ magnification. The movie is sped up by $\times 2$. 

\item Movie S7: Composite movie showing the typical evolution of the system at different cell concentrations $\rho_0$. The wavelength of the pattern depends on the cell concentration. At $\rho_0=1.2\times 10^5$cells/mL, cells migrate towards the center of the dish but don't develop any branch. The movies are sped up by $\times 40$. 

\item Movie S8: Composite movie showing the typical evolution of the system at different light intensity $I_{\rm LED}$. The wavelength of the pattern depends on the light intensity. At $I_{\rm LED}=10 {\rm mW/m^2}$, cells still mostly migrate towards the center of the dish but don't develop any branch. The movies are sped up by $\times 40$. 

\item Movie S9: Composite movie showing the typical evolution of the system for different medium viscosity $\eta$. The wavelength of the pattern depends on the viscosity. The movies are sped up by $\times 40$. 

\item Movie S10: Movie showing the evolution of the system when using the eye-deficient mutant CC1101. Cells mostly migrate towards the center of the dish but don't develop any branch. The movie is sped up by $\times 40$. 

\item Movie S11: Movie showing the typical evolution of the system at large cell concentration $\rho_0=7.5\times 10^6$cells/mL. The cells do not form a well-defined pattern anymore. The movie is sped up by $\times 40$. 

\item Movie S12: Composite movie showing the typical evolution of the system for different LED height $H_{\rm LED}$. The wavelength of the pattern depends on the height. The movies are sped up by $\times 10$. 

\item Movie S13: Movie showing the typical evolution of the system at large LED height $H_{\rm LED}=15.5$mm. Branches appear on the periphery (but not in the central region), as opposed to a prediction based on our simple 1D approach. The movie is sped up by $\times 10$. 

\end{itemize}

\section{Supplementary figures}

\renewcommand{\thefigure}{S\arabic{figure}}

\setcounter{figure}{0}

\begin{figure}
\centering
\includegraphics[width=1\textwidth]{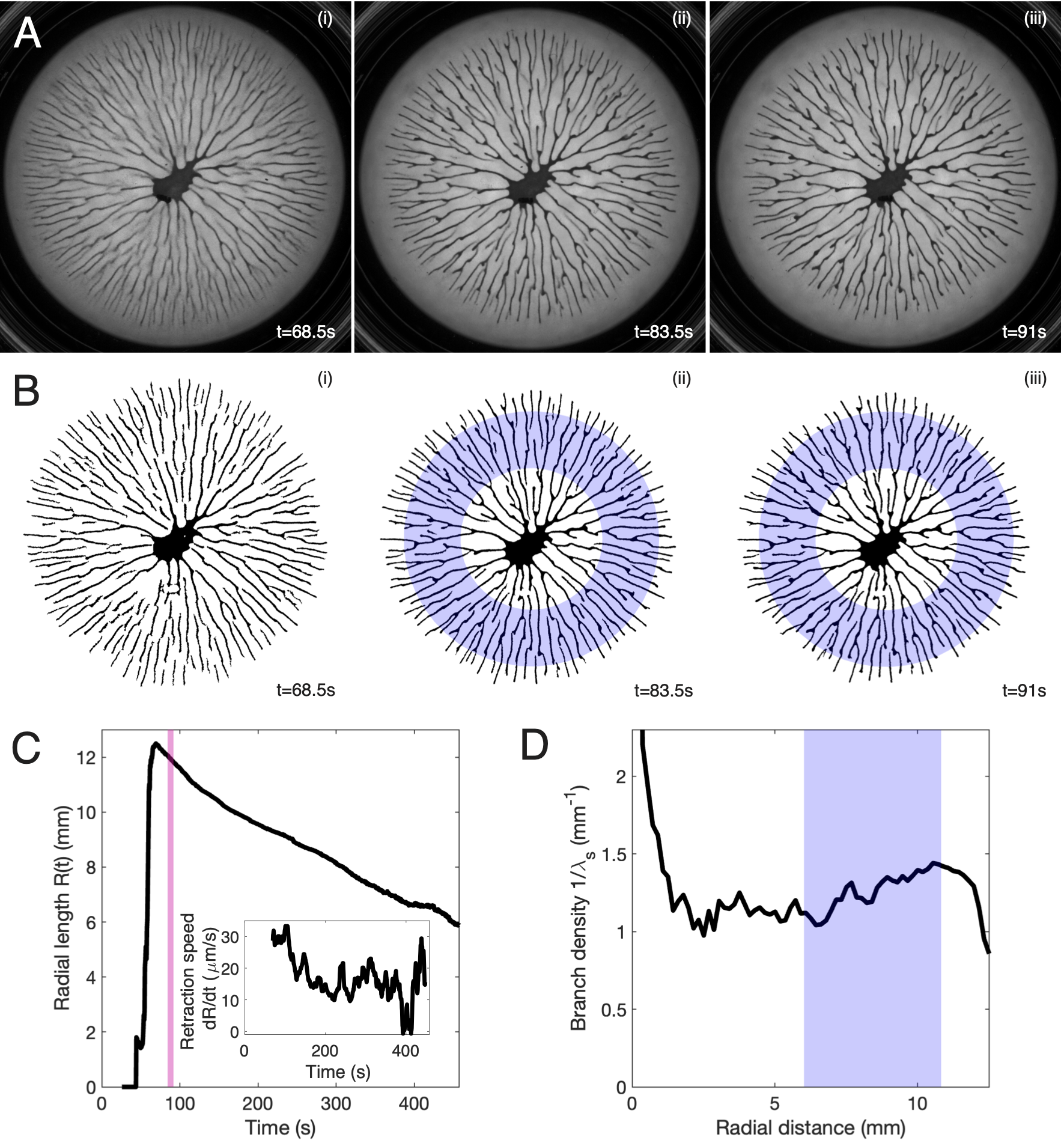}
\label{fig:6}
\caption{A: raw images and B: binarized masks (i): before onset of instability (corresponding to the maximum radial distance in panel C, (ii) and (iii), respectively first and last frame used to compute $k_s$. C. Mean radial length of the pattern $R$ versus time. $k_s$ is computed 15s to 22.5s after the peak (which corresponds to (i) in panel A and B, pattern not yet fully defined). Inset: Global retraction speed of the pattern versus time (derivative of the signal $R(t)$). D. Branch density $1/\lambda_s$ as a function of the radial distance. The branch density is roughly constant over a wide range. We then average our measurements over a ring between 50\% and 90\% of the total radial length (blue shaded area).}
\label{FigS1}
\end{figure}

\begin{figure}
\centering
\includegraphics[width=\textwidth]{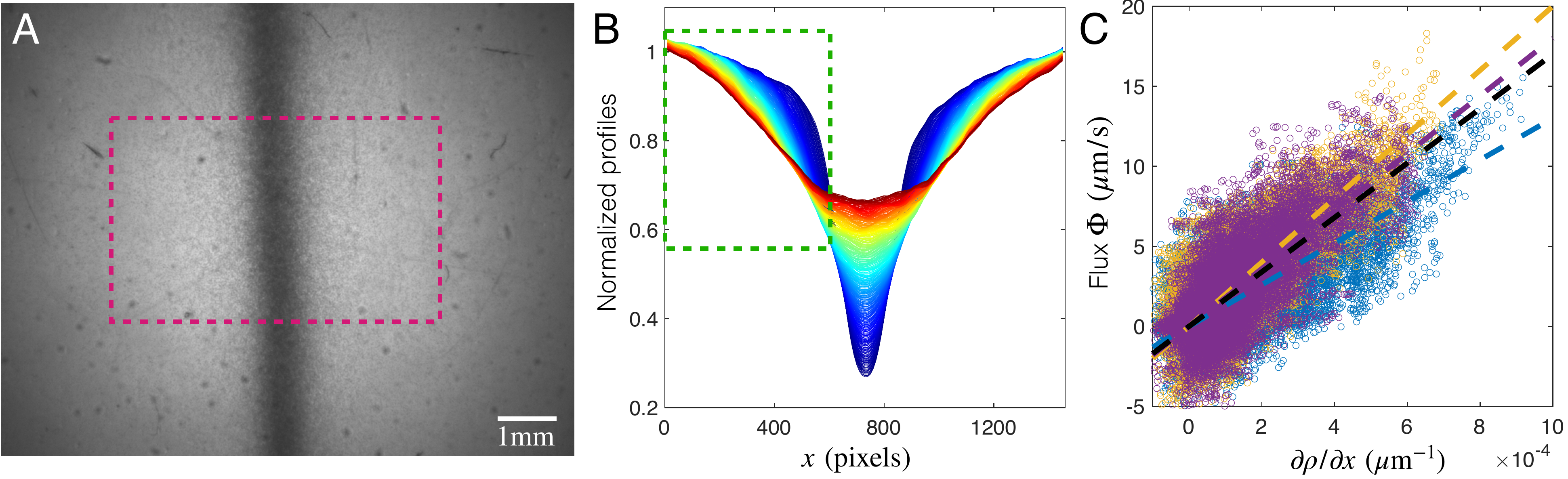}
\caption{A. Raw image of the initial configuration just after turning off the side lights. B. Density profiles in the pink region of panel A as a function of time (blue to red, total duration is 47s, profiles shown every 100ms). To extract the profiles, the raw images were divided by a background image produced from recording during 20s the homogeneous system prior to turning on the side lights. C. Gradients of cell density as a function of the flux for the 3 replicas (purple, yellow, blue), extracted in the green region shown in panel B. The data appear linearly correlated as expected. Fitted linear functions are shown in color dashed lines and the average of the 3 replicas is shown in black dashed line. We obtain a value $D_0=1.7\pm 0.4 \times 10^4 {\rm \mu m^2/s}$.}
\label{FigS2}
\end{figure}

\begin{figure}
\centering
\includegraphics[width=1\textwidth]{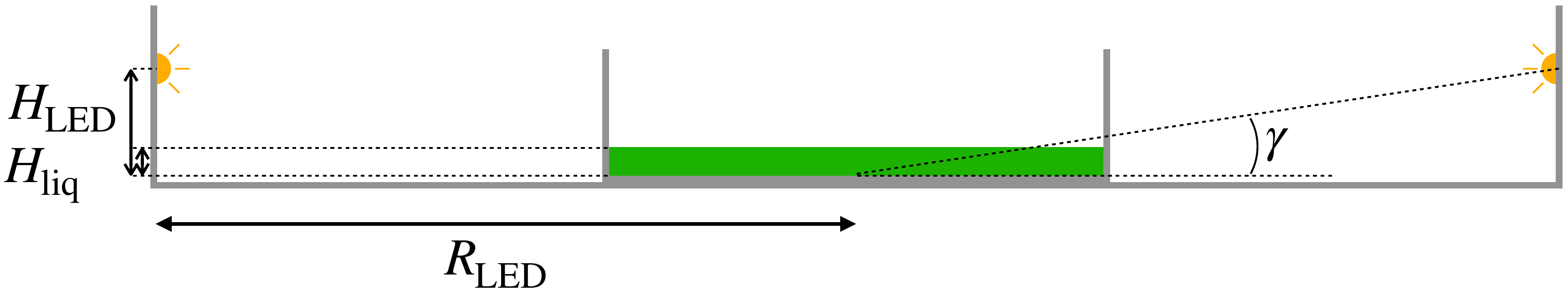}
\caption{Schematics showing a side view of the experimental setup. The drawing respects the scales of the setup. We have $H_{\rm liq}\approx 2 {\rm mm}$, $H_{\rm LED}\approx 7.5 {\rm mm}$, $R_{\rm LED}\approx 45 {\rm mm}$, leading to a typical light-incidence angle $\gamma= {\rm atan}(H_{\rm LED}/R_{\rm LED})\approx0.16 {\rm rad}\approx9.5^{\circ}$.}
\label{FigS3}
\end{figure}

\begin{figure}
\centering
\includegraphics[width=1\textwidth]{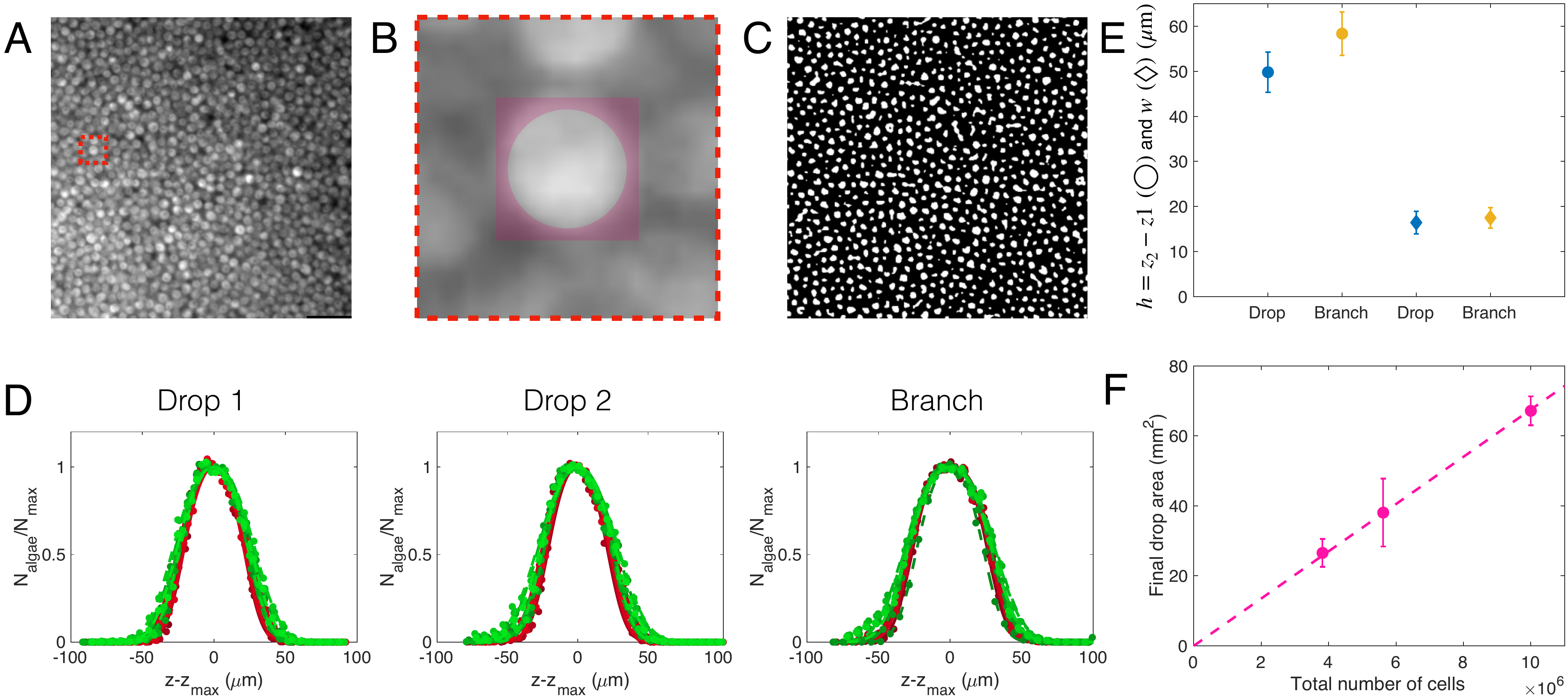}
\caption{A. Example of a raw image obtained in the bulk of a dense drop. B. Close up on the red dashed square in panel A. The pink shaded area shows the Kernel $K(r)$ used to convolute the raw images. $K$ is equal to $-1$ in the pink shaded area, while it is equal to 1 in the central disk. C. Binarized image obtained after thresholding the convoluted image. We used these binarized images to count the number of cells at each position $z$ of the vertical stack. D. Distributions of cells as a function of vertical height for two different final drops and one branch. These distributions were fitted by the function: $f(z)=N_{\rm max}/2\Big[\erf\big((z-z_1)/w\big)+\erf\big((z_2-z)/w\big)\Big]$. Experimental and fitted distributions were then normalized by the fitting parameter $N_{\rm max}$. The abscissa is $z-z_{\rm max}$ where $z_{\rm max}=(z_1+z_2)/2$ was also obtained from the fit. Green and red curve correspond to two different thresholds for the segmentation of the cells, showing no significant difference. The different curves correspond to the profiles in different regions of the images. E. Average thickness of the dense phase $h=z_2-z_1$ (circles) obtained from the fitting of the experimental distributions for both the final drop (blue) and the branches (yellow). We also show the average widths $w$ obtained from the fits (diamonds), which are consistent with the depth of focus of our imaging system. F. Final drop area as a function of the total cell number in the system. The two quantities are proportional, showing that the thickness $h$ doesn't change. From the proportionality coefficient and the thickness $h_{\rm drop}$ we can extract the density of the final dense drop.}
\label{FigS4}
\end{figure}

\begin{figure}
\centering
\includegraphics[width=0.9\textwidth]{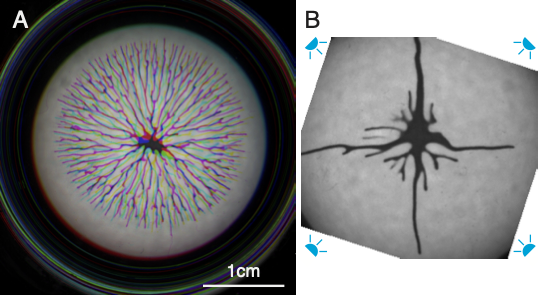}
\caption{A. Cyan-Yellow-Purple superimposition of 3 replicas obtained with the same algal culture and the same LEDs configuration (16 regularly spaced LEDs kept at the same positions), showing almost no overlap between the 3 patterns except in the center. Images taken at the same time after turning on the light sources at $\rho_0=2.8\times10^6$ cells/mL. This shows that the discreteness of the LED sources does not influence the pattern formation. B. Pattern obtained when opaque taping all the LEDs but 4. In that case, the 4 long branches don't appear because of the instability described in the main text but simply because the light field is not homogeneous.}
\label{FigS5}
\end{figure}

\begin{figure}
\centering
\includegraphics[width=0.6\textwidth]{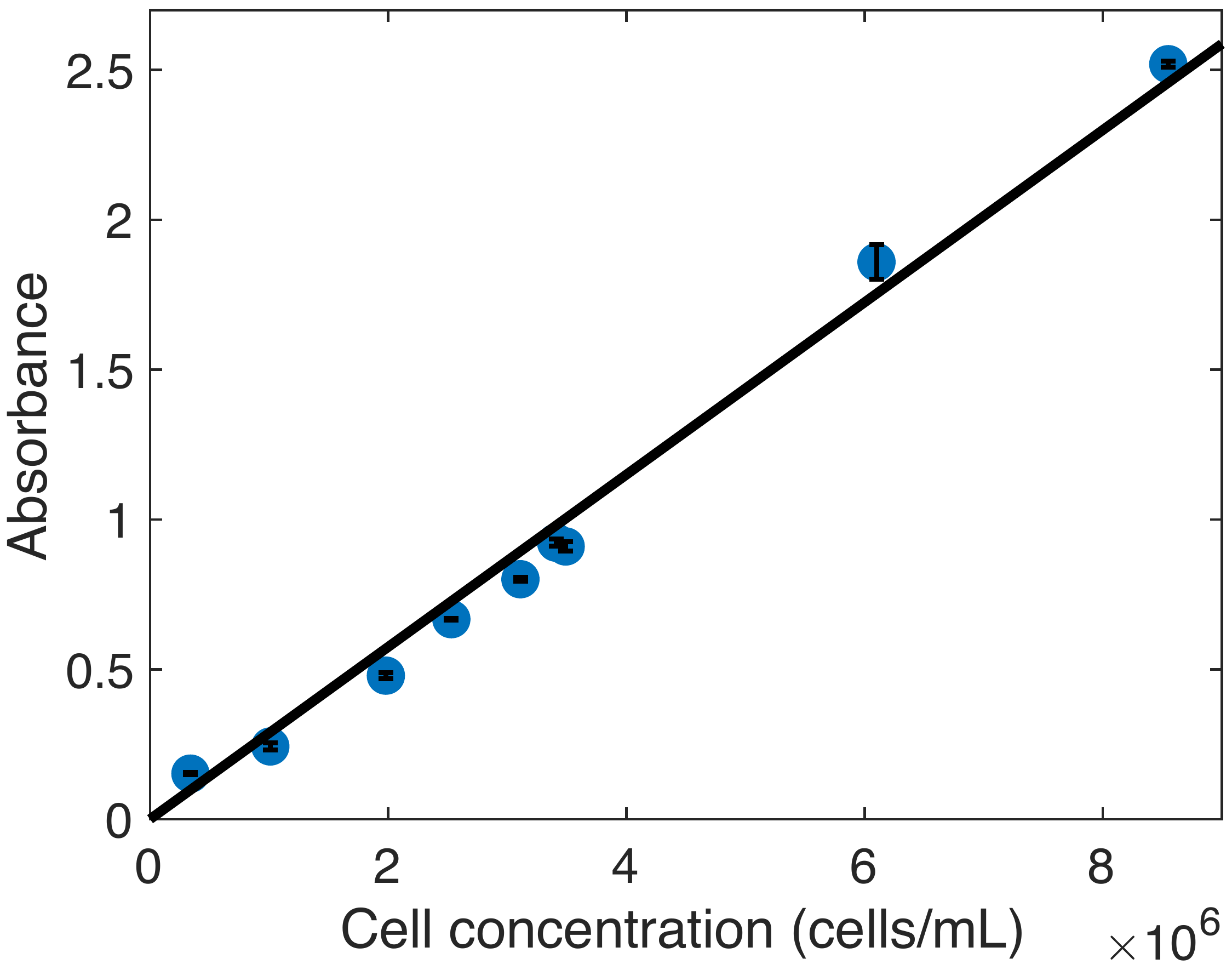}
\caption{Absorbance $A=\log(I_0/I_T)$ (at 460nm) of the cell suspension as a function of the cell concentration. The cuvette used for these measurements was $1$cm long. From a linear fit (black line) we extract the absorption coefficient per unit length and cell density of the cell suspension, $\kappa_{\rho}=3\times 10^{-7} {\rm cm^2}$.}
\label{FigS6}
\end{figure}

\begin{figure}
\centering
\includegraphics[width=0.6\textwidth]{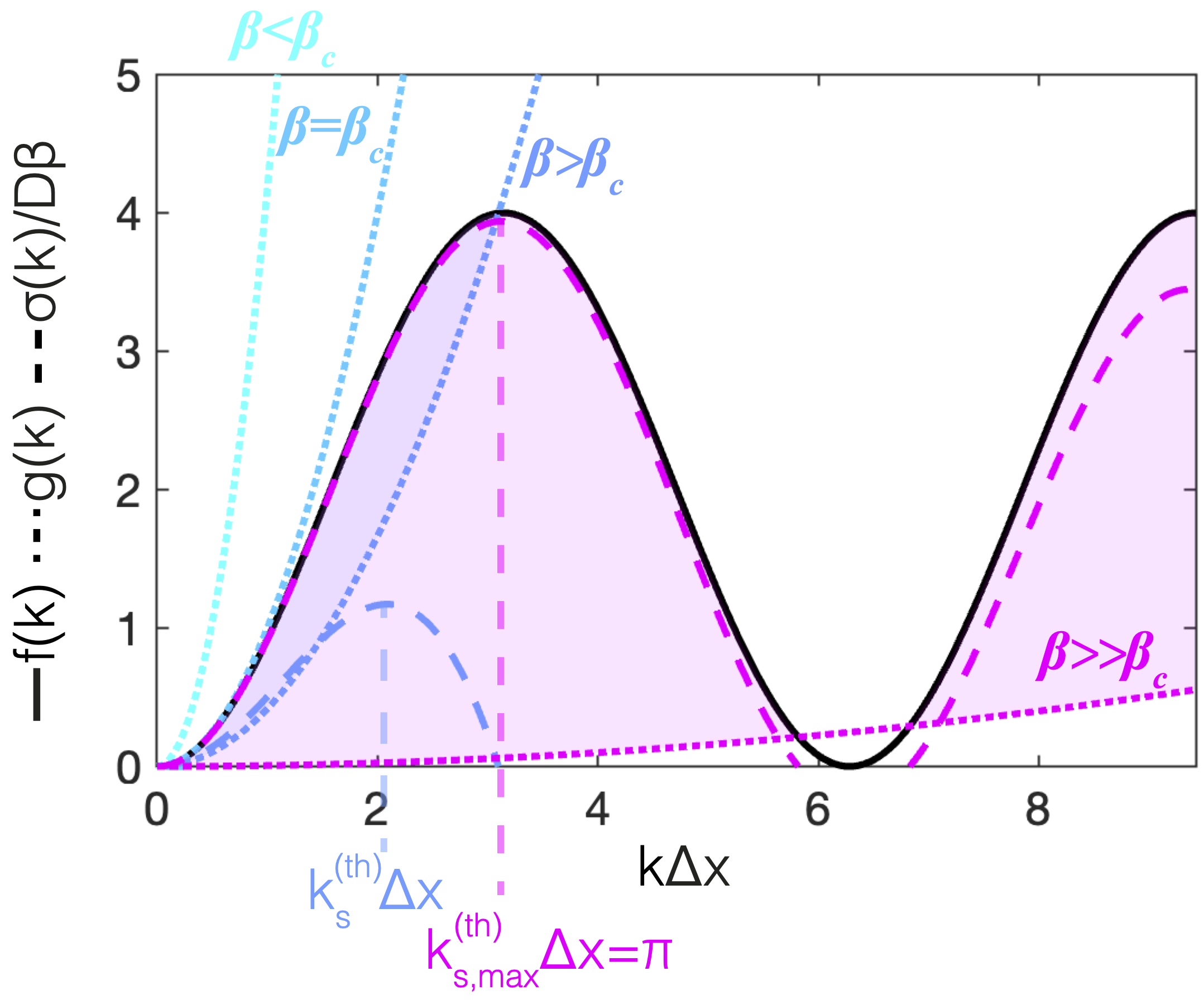}
\caption{Plot of the function $f(k)$ (solid black line) and $g(k)$ (dotted color lines) entering into the expression of the growth rate $\sigma/D\beta$ (dashed color lines) for different values of $\beta$, below (cyan), at (light blue) and above (dark blue and pink) $\beta_c=1/\Delta x^2$. Above $\beta_c$ the function $f(k)$ is larger than $g(k)$ for a range of $k$ (shaded area) and the system is unstable.  At large $\beta$ (compared to $\beta_c$, pink curve), the selected wavenumber tends to a finite value $k_{s,{\rm max}}^{\rm (th)}=\pi/\Delta x$ (pink dashed line).}
\label{FigS6bis}
\end{figure}


\begin{figure}
\centering
\includegraphics[width=1\textwidth]{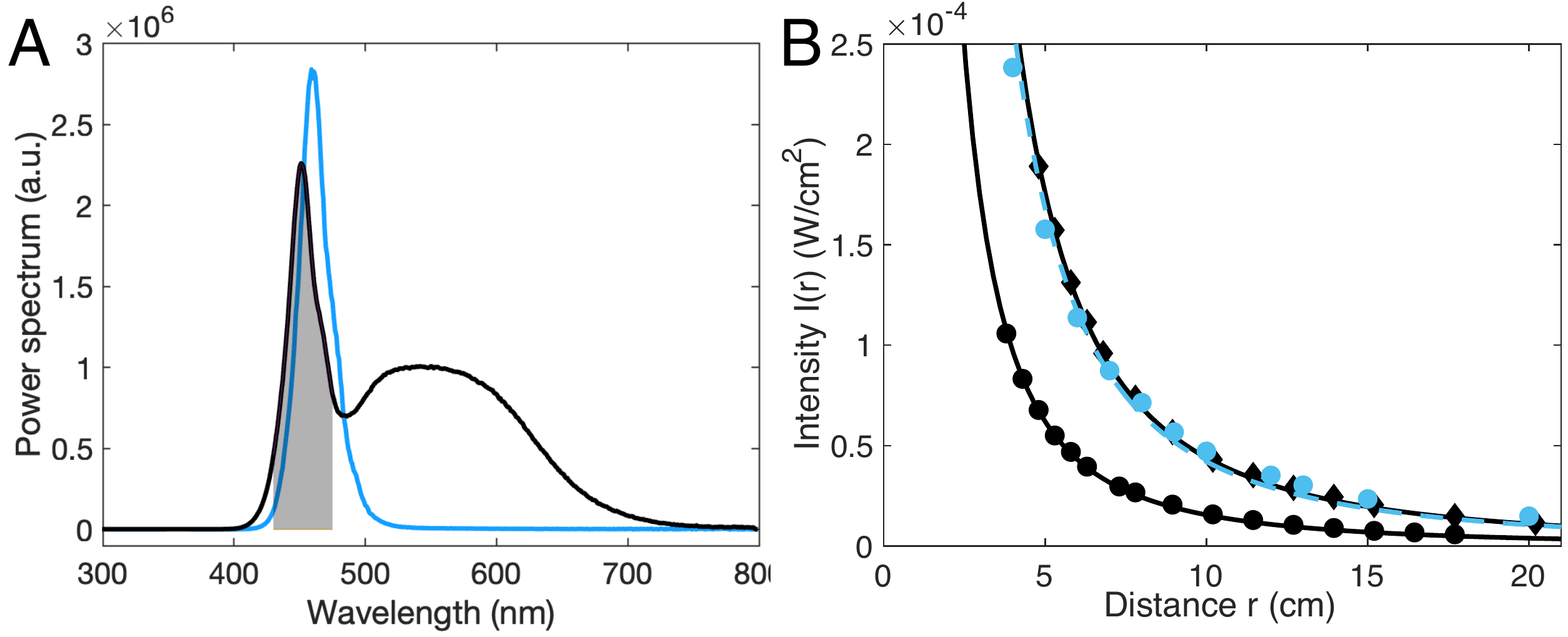}
\caption{A. Power spectrum of the white and blue LED (black and blue lines respectively) at applied voltage $U=10$V. The black shaded area shows the area of the spectrum in the range $[430$-$475]$nm used to obtain the total light power of our white LED. B. To that purpose we measured the light intensity at several distances $r$ of the white LED in the range $[430$-$475]$nm at the two voltages $U_{\rm low}=10$V (circles) and $U_{\rm high}=14.6$V (diamonds) and fitted by $I(r)=P_{430/475}/r^2$ (black solid lines) to extract the emitted light power in this range of wavelength. From measuring the area under the power spectrum in this range of wavelength (black shading in panel A), we can then estimate the total emitted light power at any applied voltage $U$ by simply measuring the total area under the spectrum at that voltage. The blue curve in panel B is the measured intensity of the blue LED used in the experiments to quantify photo-protection as a function of the distance which was fitted by $I(r)=P_{\rm blue}/r^2$ (dashed line) to compute the intensity $I_{\rm tot}=78 {\rm W/m^2}$.}
\label{FigS8}
\end{figure}

\begin{figure}
\centering
\includegraphics[width=1\textwidth]{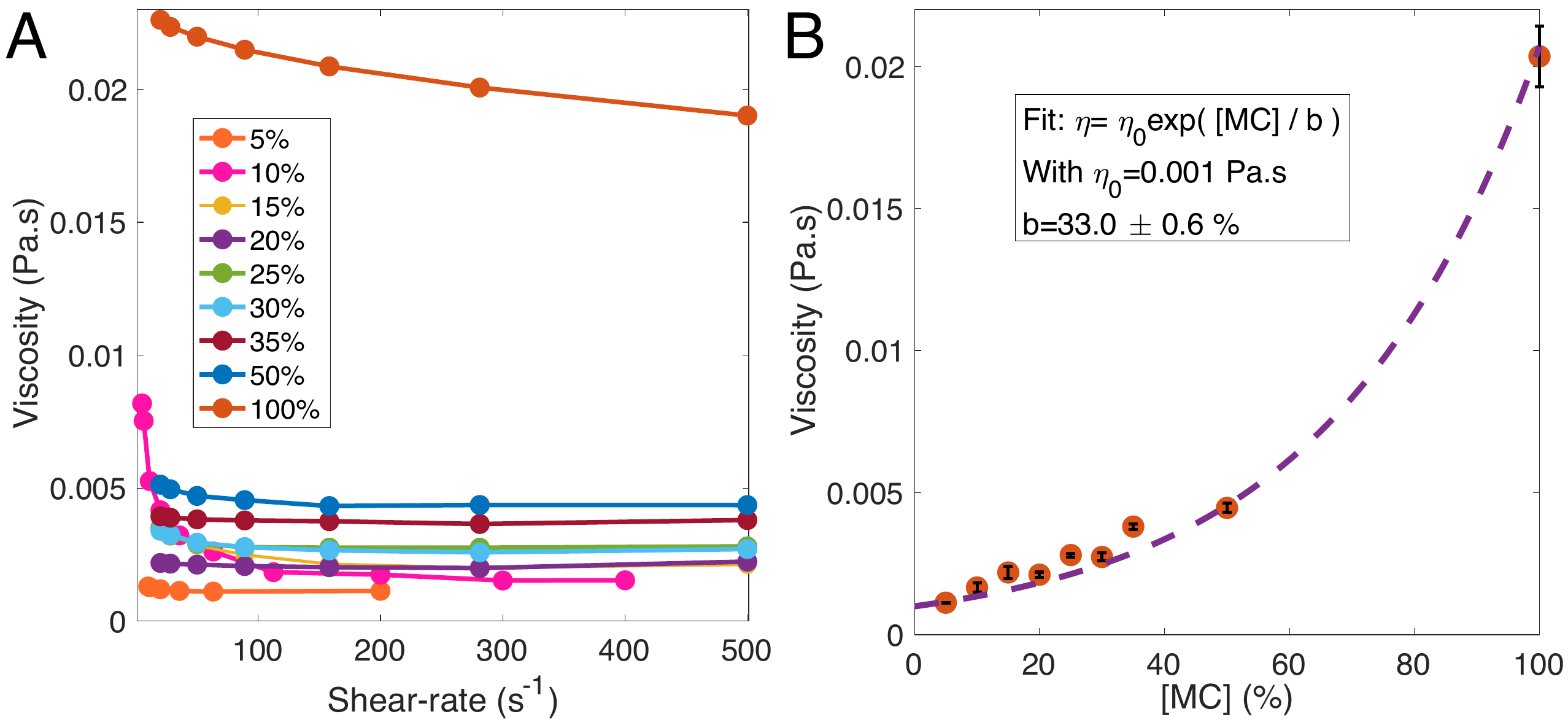}
\caption{A. Measured viscosity as a function of imposed shear rate for different solutions of Methylcellulose. B. High shear rate viscosity as a function of the concentration of Methylcellulose (expressed as a percentage of the mother MC solution added, see text section S8).}
\label{FigS9}
\end{figure} 

\begin{figure}
\centering
\includegraphics[width=1\textwidth]{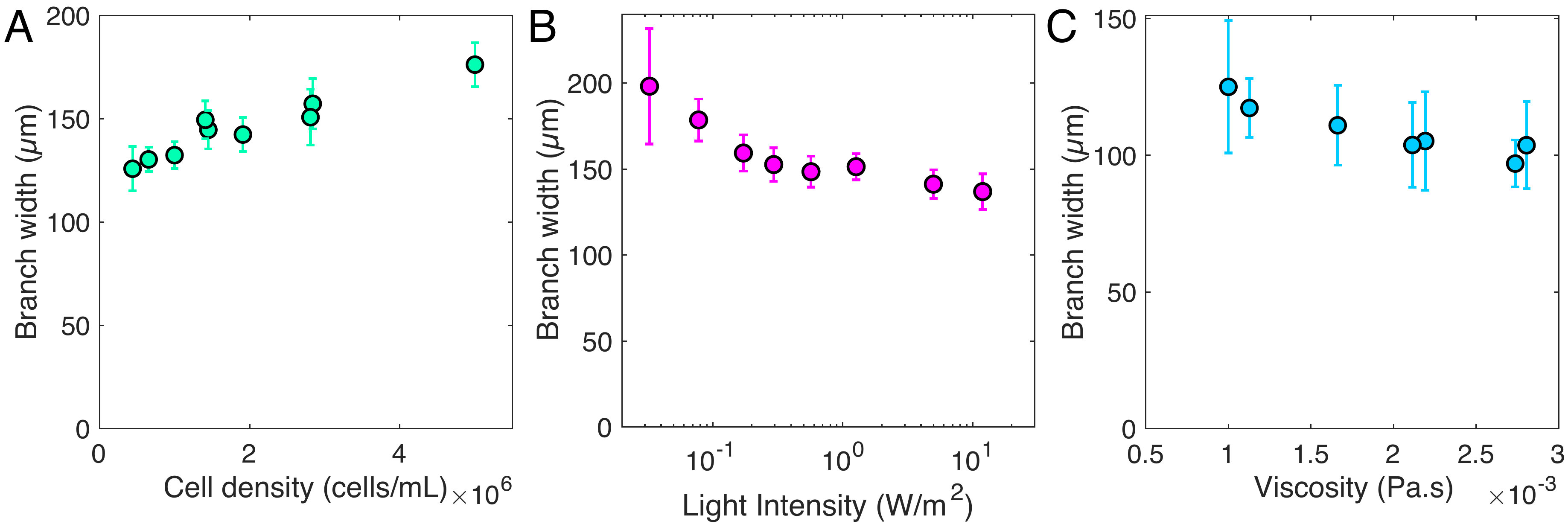}
\caption{Average branch width (measured at the same times as $k_s$, see section S1) for the different experimental parameters that were varied, A. initial cell density $\rho_0$, B. light intensity $I_{\rm LED}$ and C. medium viscosity $\eta$.}
\label{FigS10}
\end{figure} 

\begin{figure}
\centering
\includegraphics[width=0.5\textwidth]{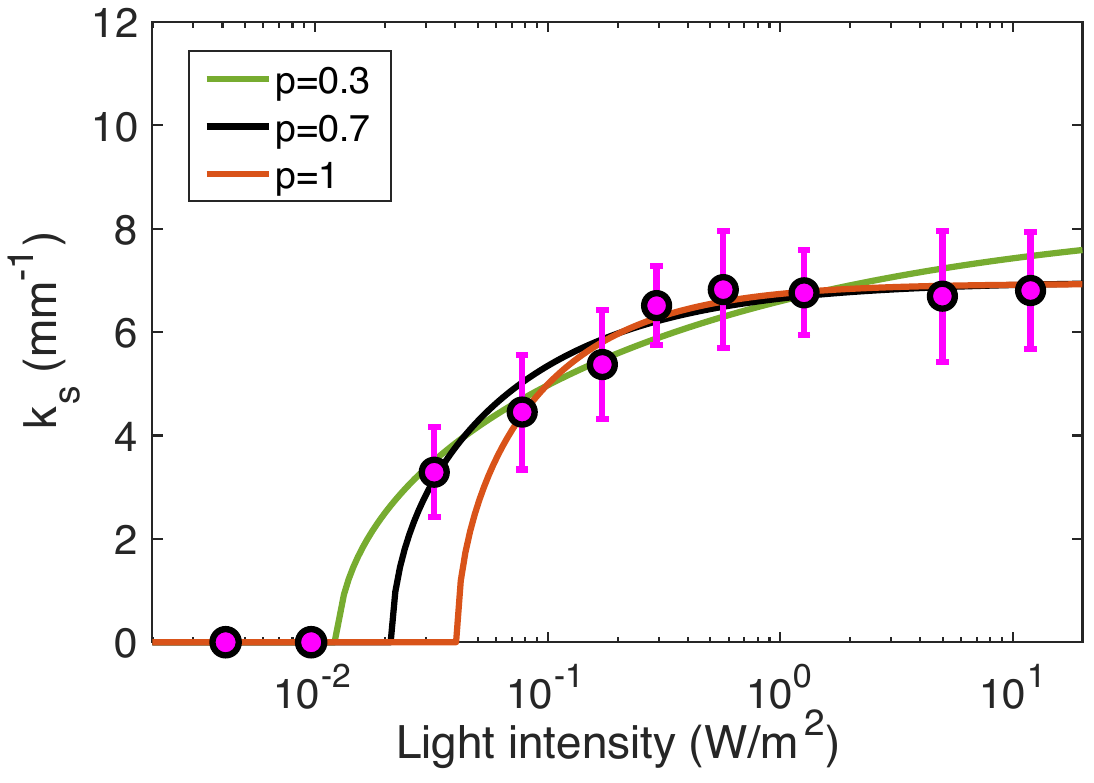}
\caption{Best fits obtained on the dataset "intensity" when fixing $p$ to different values and leaving $K$ and $\nu$ as free fitting parameters (still fixing also $\Delta x$ to the average value from the "density" and "viscosity" experiments). Values close to $p=0.7$ gave the best fits. If too low the high intensity saturation is not well captured, and if too high the low intensity threshold is not well determined. In this analysis $p$ was varied between 0.2 and 2. }
\label{FigS11}
\end{figure} 

\begin{figure}
\centering
\includegraphics[width=1\textwidth]{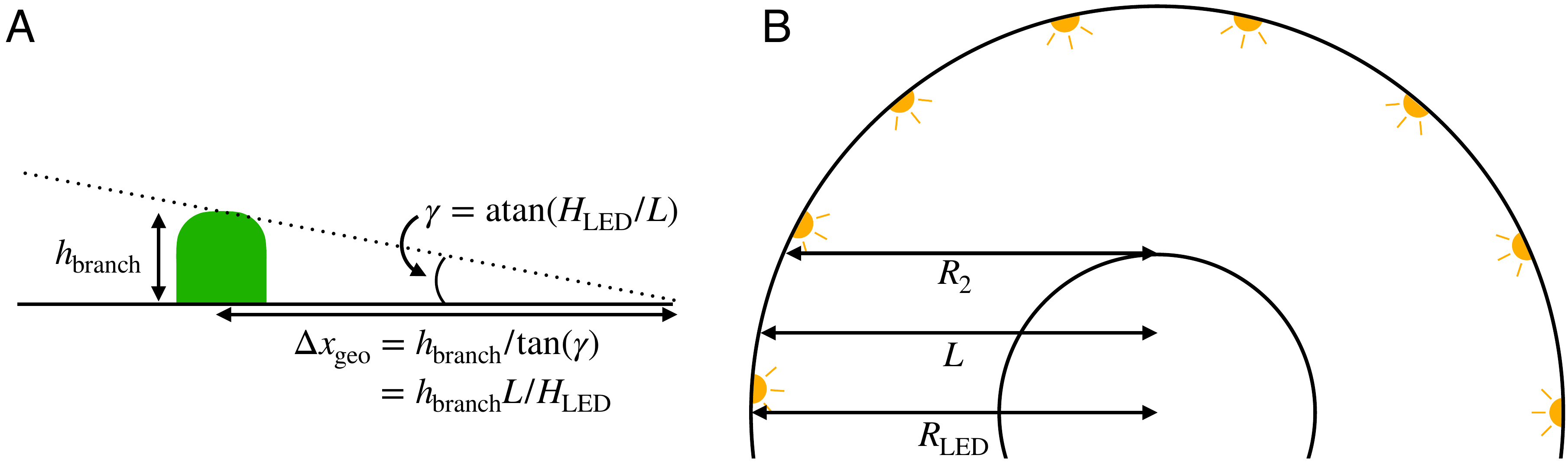}
\caption{A. Schematics showing the definition of $\Delta x_{\rm geo}(H_{\rm LED})$. B. Schematics showing the definition of the length $L$, that was varied between $R_{\rm LED}$ and $R_2$ for plotting the theoretical curves in Fig. 3F.}
\label{FigS11b}
\end{figure} 

\begin{figure}
\centering
\includegraphics[width=1\textwidth]{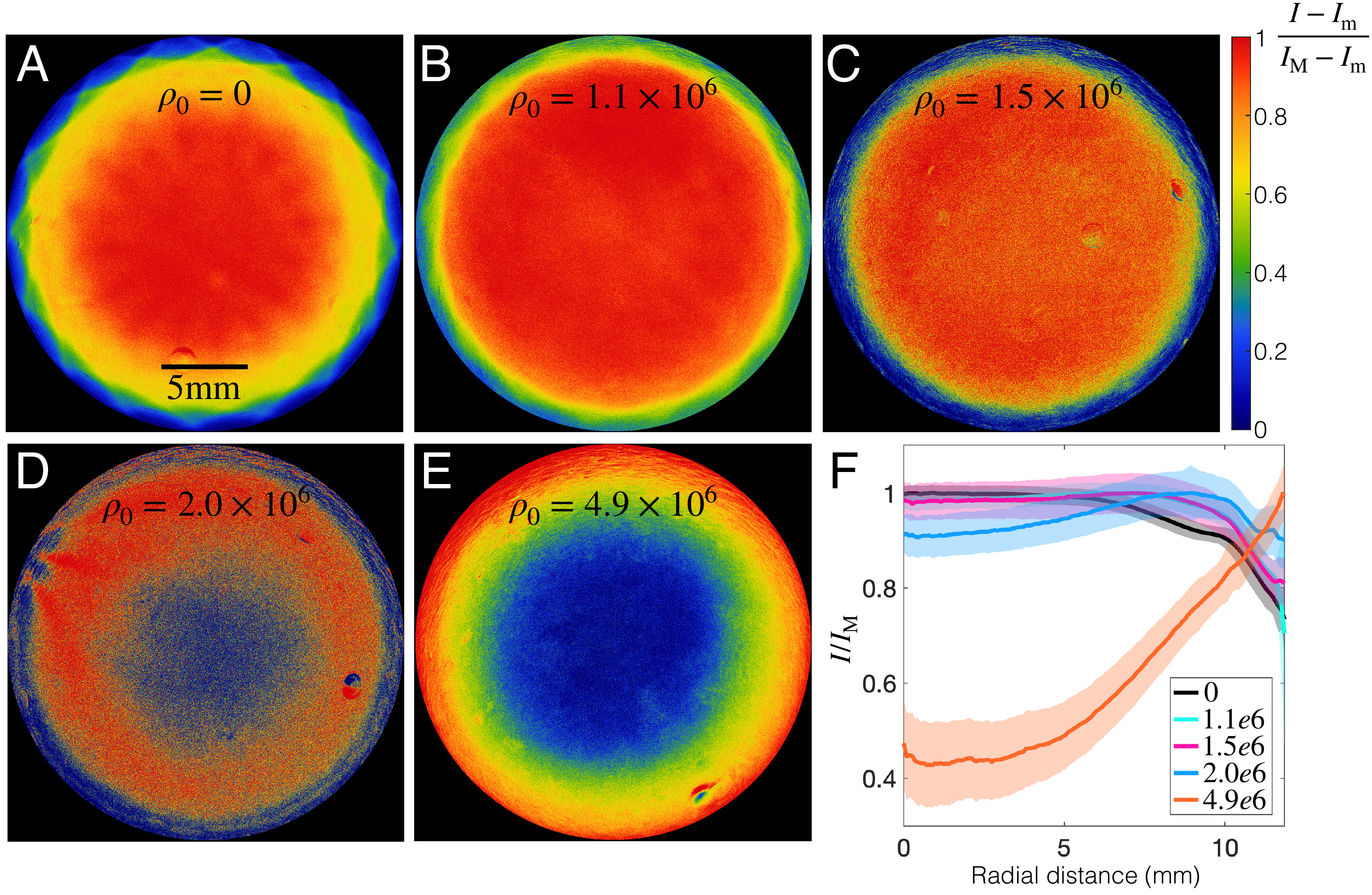}
\caption{A-E. Measured intensity fields (from the sixteen LEDs of the setup) using a recently developed fluorescence technique \cite{Lahlou2023} for different concentrations of cells representative of the branch instability experiments. Here we show the normalized intensity $(I-I_{\rm m})/(I_ {\rm M}-I_{\rm m})$ with $I_{\rm m}$ and $I_{\rm M}$ the minimal and maximal intensities of the field respectively, in order to better observe the spatial variations. F. Radial intensity profile for all cases shown in panels A-E. Here the intensity has been simply normalized by the maximal intensity of the field, in order to better show the relative variations for each cell concentration.} 
\label{FigS12}
\end{figure}

\end{document}